\RequirePackage[mathlines]{lineno}
\documentclass[twocolumn,showpacs,aps,prl]{revtex4-2}
\usepackage{overpic,graphicx}
\usepackage{dcolumn}
\usepackage{bm}
\usepackage{rotating}
\usepackage{subfigure}
\usepackage{color}
\usepackage[bookmarksnumbered, pdfstartview=FitH,colorlinks,urlcolor=blue, citecolor=blue,linkcolor=blue,] {hyperref}
\usepackage{amsmath}
\usepackage{lineno}
\usepackage{multirow}\usepackage{csquotes}
\usepackage{makecell}%2020.12.10, make the text to be center/right/left
\usepackage{etoolbox}
\makeatletter
\patchcmd{\doauthor}{\unskip}{}{}{}
\makeatother
\newcommand{\psip}{\psi(3686)}

\let\oldequation\equation
\let\oldendequation\endequation
\renewenvironment{equation}
{\linenomathNonumbers\oldequation}
{\oldendequation\endlinenomath}

\newcommand{\BESIIIorcid}[1]{\href{https://orcid.org/#1}{\hspace*{0.1em}\raisebox{-0.45ex}{\includegraphics[width=1em]{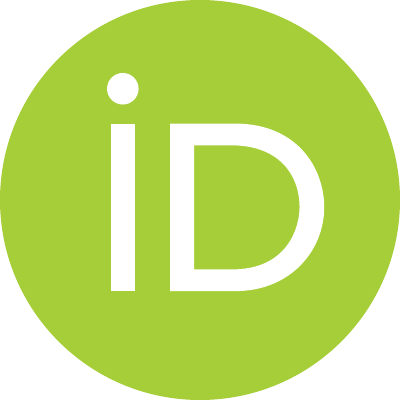}}}} 
\begin{document}

\let\oldequationstar\equation
\let\oldendequationstar\endequation
\renewenvironment{equation*}
{\linenomathNonumbers\oldequationstar\nonumber}
{\oldendequationstar\endlinenomath}

\hyphenpenalty=10000
\tolerance=1000
%\linenumbers

\title{\bf \boldmath
First Measurement of the Absolute Branching Fraction of $\eta_c \to \gamma \gamma$
}

\author{%% Saved at => 2025-08-28
M.~Ablikim$^{1}$\BESIIIorcid{0000-0002-3935-619X},
M.~N.~Achasov$^{4,c}$\BESIIIorcid{0000-0002-9400-8622},
P.~Adlarson$^{81}$\BESIIIorcid{0000-0001-6280-3851},
X.~C.~Ai$^{86}$\BESIIIorcid{0000-0003-3856-2415},
R.~Aliberti$^{39}$\BESIIIorcid{0000-0003-3500-4012},
A.~Amoroso$^{80A,80C}$\BESIIIorcid{0000-0002-3095-8610},
Q.~An$^{77,64,\dagger}$,
Y.~Bai$^{62}$\BESIIIorcid{0000-0001-6593-5665},
O.~Bakina$^{40}$\BESIIIorcid{0009-0005-0719-7461},
Y.~Ban$^{50,h}$\BESIIIorcid{0000-0002-1912-0374}
% ...
\\
\vspace{0.2cm}
(BESIII Collaboration)\\
\vspace{0.2cm} {\it
$^{1}$ Institute of High Energy Physics, Beijing 100049, People's Republic of China\\
% ...
$^{4}$ Budker Institute of Nuclear Physics SB RAS (BINP), Novosibirsk 630090, Russia\\
% ...
$^{39}$ Johannes Gutenberg University of Mainz, Johann-Joachim-Becher-Weg 45, D-55099 Mainz, Germany\\
$^{40}$ Joint Institute for Nuclear Research, 141980 Dubna, Moscow region, Russia\\
% ...
$^{50}$ Peking University, Beijing 100871, People's Republic of China\\
% ...
$^{62}$ Southeast University, Nanjing 211100, People's Republic of China\\
% ...
$^{64}$ State Key Laboratory of Particle Detection and Electronics, Beijing 100049, Hefei 230026, People's Republic of China\\
% ...
$^{77}$ University of Science and Technology of China, Hefei 230026, People's Republic of China\\
% ...
$^{80}$ University of Turin and INFN, (A)University of Turin, I-10125, Turin, Italy; (B)University of Eastern Piedmont, I-15121, Alessandria, Italy; (C)INFN, I-10125, Turin, Italy\\
$^{81}$ Uppsala University, Box 516, SE-75120 Uppsala, Sweden\\
% ...
$^{86}$ Zhengzhou University, Zhengzhou 450001, People's Republic of China\\

\vspace{0.2cm}
$^{\dagger}$ Deceased\\
% ...
$^{c}$ Also at the Novosibirsk State University, Novosibirsk, 630090, Russia\\
% ...
$^{h}$ Also at State Key Laboratory of Nuclear Physics and Technology, Peking University, Beijing 100871, People's Republic of China\\
% ...

}
%% ends here %%
}

\begin{abstract} 
We apply a tag-and-probe method to precisely measure the absolute branching fraction of the decay $\eta_c \to \gamma \gamma$ with the BESIII experiment at BEPCII. Starting with a large initial sample of $2712.4\pm 14.3$ million $\psi(3686)$ events, a sample of 0.16 million $\eta_c$ events are tagged via the golden channel $\psi(3686)\to \pi^0 h_c$, $h_c\to \gamma \eta_c$, effectively avoiding interference effects. The absolute branching fraction of $\eta_c \to \gamma \gamma$ is measured for the first time to be $\mathcal{B}(\eta_c \to \gamma \gamma) = (2.45 \pm 0.48_{\rm stat.} \pm 0.09_{\rm syst.}) \times 10^{-4}$. Using the world average value of the total width of the $\eta_c$, the partial decay width of $\eta_c \to \gamma \gamma$ is calculated to be $\Gamma(\eta_c \to \gamma \gamma) = (7.48 \pm 1.48_{\rm stat.} \pm 0.30_{\rm syst.})~{\rm keV}$.

\end{abstract}

%\pacs{13.75.Ev}

\maketitle

\oddsidemargin  -0.2cm
\evensidemargin -0.2cm

%%%%%%%%%%%%%%%%%%%%%%%%%% Introduction %%%%%%%%%%%%%%%%%%%%%%%%%%%%%

As the simplest decay mode of the ground state of charmonium, the decay $\eta_c \to \gamma \gamma$ plays a special role as a benchmark to test theoretical models. The electromagnetic annihilation of the $c\bar{c}$ pair provides access to the decay constant and the wave function of the $\eta_c$. Compared to hadronic decays of the $\eta_c$, the decay $\eta_c \to \gamma \gamma$ provides a much cleaner environment in which to treat the non-perturbative effects that arise in $c\bar{c}$ annihilation.  Due to the complexity of these non-perturbative effects, considerable effort has historically been devoted to calculations of the $\eta_c$ two-photon decay width using methods that include non-relativistic quantum chromodynamics (QCD)~\cite{Czarnecki:2001zc, Bodwin:2002cfe, Brambilla:2006ph, Guo:2011tz, Jia:2011ah, Feng:2017hlu, Brambilla:2018tyu, Yu:2019mce, Abreu:2022cco}, the relativistic quark model~\cite{Ackleh:1991dy, Munz:1996hb, Ebert:2003mu, Kim:2004rz}, Schwinger-Dyson and Bethe-Salpeter equations~\cite{Chao:1996bj, Chen:2016bpj, Higuera-Angulo:2024oui}, sum rules~\cite{PhysRevD.79.074026}, heavy quarkonia spin symmetry~\cite{Lansberg:2006dw}, soft-gluon factorization~\cite{Li:2019ncs}. Recently, \textit{ab initio} calculations with unprecedented accuracy have been provided by lattice QCD~\cite{Dudek:2006ut, CLQCD:2016ugl, CLQCD:2020njc, Liu:2020qfz, Zhang:2021xvl, Colquhoun:2023zbc, Meng:2021ecs, Colquhoun:2023zbc}, among which the most precise result so far is from the HPQCD group, finding $\Gamma(\eta_c \to \gamma \gamma) = (6.788 \pm 0.045_{\rm fit} \pm 0.041_{\rm syst.})~{\rm keV}$~\cite{Colquhoun:2023zbc}.

At the same time, non-perturbative uncertainties can be cancelled by forming ratios of various observables, allowing for clean perturbative QCD predictions~\cite{Czarnecki:2001zc}.  To leading order in $\alpha_{s}$, the ratio of the decay rates for the simplest decays of the lowest-lying charmonia states, 
\begin{equation}
	\label{equ:R}
	\mathcal{R} \equiv \frac{\Gamma(J/\psi \to e^+ e^-)}{\Gamma(\eta_c \to \gamma \gamma)},
\end{equation}
is predicted to be about 3/4, and is free from any non-perturbative uncertainties.

Experimental measurements, which serve to examine the theoretical predictions, have, in contrast, long been technically difficult. The first explorations of the $\eta_c$ two-photon width began using the two-photon fusion process $\gamma^{(*)} \gamma^{(*)} \to \eta_c$ with the $\eta_c$ decaying to hadronic final states~\cite{ParticleDataGroup:2024cfk}.  This method introduces uncertainties that arise from the form factor of the virtual photon and the branching fractions of $\eta_c$ decays. Subsequently, the branching fraction of $\eta_c \to \gamma \gamma$ was measured using the decay $B^+ \to \eta_c K^+$ at Belle~\cite{Belle:2006mzv} and the process $p \bar{p} \to \eta_c$ in $p \bar{p}$ collisions~\cite{AnnecyLAPP-CERN-Genoa-Lyon-Oslo-Rome-Strasbourg-Turin:1987qkd, E760:1995rep, FermilabE835:2003ula}. In 2008, CLEO reported the first hint of $\eta_c \to \gamma \gamma$ via the radiative decay $J/\psi \to \gamma \eta_c$~\cite{CLEO:2008qfy, BESIII:2012lxx, BESIII:2024rex}. The product branching fraction was measured to be $\mathcal{B}(J/\psi \to \gamma \eta_c) \times \mathcal{B}(\eta_c \to \gamma \gamma) = (1.2^{+2.7}_{-1.1} \pm 0.3) \times 10^{-6}$. Following a similar method, BESIII reported the same product branching fraction in 2013~\cite{BESIII:2012lxx} and updated it using a larger data sample in 2025~\cite{BESIII:2024rex}, obtaining $\mathcal{B}(J/\psi \to \gamma \eta_c) \times \mathcal{B}(\eta_c \to \gamma \gamma) = (4.5 \pm 1.2 \pm 0.6) \times 10^{-6}$ and $(5.23 \pm 0.26 \pm 0.30) \times 10^{-6}$, respectively. Strong interference effects complicate the radiative decays of the $J/\psi$ or $\psi(3686)$ to an $\eta_c$ at electron-positron colliders. In addition, all of the previous measurements involve product branching fractions.

In this Letter, a tag-and-probe method is introduced that allows for a determination of absolute $\eta_c$ branching fractions at BESIII. The \textit{golden channel} $\psi(3686)\to \pi^0 h_c,h_c\to\gamma\eta_c$ that has been exploited in a previous BESIII analysis~\cite{BESIII:2012urf} is used to produce the $\eta_c$. This channel effectively avoids complications from interference and significantly suppresses background, a feature that was not fully exploitable in earlier studies due to limited statistics. By tagging the $\eta_c$ in a larger sample of $(2712.4\pm 14.3) \times 10^6$ $\psi(3686)$ events~\cite{BESIII:2017tvm, BESIII:2024lks} collected with the BESIII detector at the BEPCII collider, we measure the absolute branching fraction of $\eta_c \to \gamma \gamma$ for the first time.

%%%%%%%%%%%%%%%%%%%%%%%%%% BESIII %%%%%%%%%%%%%%%%%%%%%%%%%%%%%%%%%%%

The BESIII detector is a magnetic spectrometer~\cite{BESIII:2009fln} located at the Beijing Electron Positron Collider (BEPCII)~\cite{Yu:2016cof}. The cylindrical core of the BESIII detector consists of a helium-based multilayer drift chamber (MDC), a plastic scintillator time-of-flight system (TOF), and a CsI(Tl) electromagnetic calorimeter (EMC), which are all enclosed in a superconducting solenoidal magnet providing a 1.0~T magnetic field.

To obtain the absolute branching fraction, the analysis is divided into a \textit{Tag Side} and a \textit{Signal Side}. For the \textit{tag side}, we only reconstruct $\pi^0 \gamma$ combinations to identify the decay $\psi(3686) \to \pi^0 h_c,~h_c \to \gamma \eta_c$. We use the recoiling system to obtain the total efficiency-corrected yield of $\eta_c$ inclusive decays, denoted as $N_{\eta_c}^{\rm tag}= N^{\rm tag}_{\rm rec} / \varepsilon^{\rm tag}_{\rm rec}$. For the \textit{signal side}, we also use the two photons from the $\eta_c$ decay to get the efficiency-corrected yield of exclusive decays, $\eta_c \to \gamma \gamma$, denoted as $N_{\eta_c}^{\rm sig} = N^{\rm sig}_{\rm rec} / \varepsilon^{\rm sig}_{\rm rec}$. Here, $N^{\rm tag/sig}_{\rm rec}$ and $\varepsilon^{\rm tag/sig}_{\rm rec}$ represent the reconstructed yield and efficiency for the \textit{tag} or \textit{signal side}, respectively. The absolute branching fraction of the decay $\eta_c \to \gamma\gamma $ can be calculated from
\begin{equation}
\label{equ:branching_fraction}
\mathcal{B}(\eta_c \to \gamma \gamma) = \frac{N_{\eta_c}^{\rm sig}}{N_{\eta_c}^{\rm tag}} =\frac{N^{\rm sig}_{\rm rec} / \varepsilon^{\rm sig}_{\rm rec}}{N^{\rm tag}_{\rm rec} / \varepsilon^{\rm tag}_{\rm rec}}.
\end{equation}

Using a {\sc geant4}-based~\cite{GEANT4:2002zbu, Allison:2006ve, Allison:2016lfl} software package~\cite{2006ObjectorientedBD}, Monte-Carlo (MC) simulated data samples are produced incorporating the geometric description of the BESIII detector~\cite{Ya-Jun:2008mrr, Liang:2009zzb, Huang:2022wuo} and the detector response. An inclusive MC sample containing 2747 million $\psi(3686)$ events is used to investigate the potential backgrounds. In addition, a semi-exclusive MC sample with 24 million events of $\psi(3686) \to \pi^0 h_c,~h_c \to \gamma \eta_c$ is generated to determine the signal shape and signal efficiency for the \textit{tag side}, where the $\eta_c$ is allowed to decay inclusively. An exclusive MC sample containing 1.19 million events of $\psi(3686) \to \pi^0 h_c,~h_c \to \gamma \eta_c,~\eta_c \to \gamma \gamma$ is produced for the signal shape and the signal efficiency for the \textit{signal side}.

%%%%%%%%%%%%%%%%%%%%%%%%%%%%% tag side %%%%%%%%%%%%%%%%%%%%%%%%%%%%%%%%%%%
The reconstructed particles from the \textit{tag side} are the $\pi^0$ and $\gamma$.  Both are reconstructed from the showers collected in the EMC. The minimum energy deposited in the EMC is required to be larger than $25~{\rm MeV}$ and $50~{\rm MeV}$ for the barrel region ($|{\rm cos}\theta|<0.8$) and endcap region ($0.86<|{\rm cos}\theta|<0.92$), respectively, where $\theta$ refers to the polar angle of a certain shower. To further suppress the electronic noise and other showers unrelated to the event, the EMC time of the selected shower ($T$) must be within $[0, 700]~{\rm ns}$ after the event start time if charged tracks are reconstructed in the event. In the case of no charged tracks, we require $|T-T_0| \leq 500~{\rm ns}$, where $T_0$ refers to the time of the most energetic EMC shower. Moreover, the angle between the selected shower and its nearest charged track is required to be larger than $10^{\circ}$.

The $\pi^0$ candidates are reconstructed by combining two showers with some further requirements. Both of the two showers are required to be in the barrel region ($|{\rm cos}\theta|<0.8$) to suppress the higher background in the endcap region induced by the beam noise. The energy of each shower is also required to be larger than $40~{\rm MeV}$. Then the two showers, for which the $\gamma\gamma$ invariant mass $M(\gamma\gamma)$ must fall within $[0.120, 0.145]~{\rm GeV}/c^2$, are further selected by a kinematic fit that constrains their invariant mass to the world average value of the $\pi^0$ mass~\cite{ParticleDataGroup:2024cfk}.  The resulting residual sum of squares is required to be $\chi_{\rm 1C}^{2}<5$. This criterion is optimized using the figure of merit $S/\sqrt{S+B}$, where $S$ and $B$ represent the numbers of signal and background events, respectively, in the $h_c$ signal region  
$RM(\pi^0) \in [3.52,3.53]~{\rm GeV}/c^2$. The variable $RM(\pi^0)$ is defined as
$RM(\pi^0) = \sqrt{({p}^\mu_{e^+ e^-} - {p}^\mu_{\pi^0})^{2}}$, where ${p}^\mu_{e^+ e^-}$ and ${p}^\mu_{\pi^0}$ denote the four-momenta of the $e^+ e^-$ in the collision and the two showers of the $\pi^0$ after the kinematic fit. Meanwhile, candidate events falling inside the mass window of 
the $h_c$, defined as $RM(\pi^0) \in[3.48,3.57]~{\rm GeV}/c^2$, are retained for the further analysis. 
If multiple $\pi^0$ candidates are found, we keep the one with the smallest $\chi_{\rm 1C}^{2}$ value. 

The reconstructed $\pi^0$ candidates are then combined with other bachelor photons, which are further selected by identifying the photon from the electric dipole (E1) transition $h_c \to \gamma \eta_c$. First, the bachelor photon is excluded from being assigned as the E1 photon if it can be paired with any other photon to reconstruct a $\pi^0$ candidate that meets the following criteria: the invariant mass falls within $[0.115,0.150]~{\rm GeV}/c^2$ and $\chi^2_{\rm 1C}<200$. Considering that the energy of the E1 photon is expected to peak around 500 MeV in the center-of-mass system of the $h_c$ and taking the width $\Gamma_{\eta_c} \approx 30~{\rm MeV}$~\cite{Colquhoun:2023zbc} into account, the energy of the bachelor photon in the recoiling system of the $\pi^0$ is required to be within $[0.46,0.58]~{\rm GeV}$. If there are still multiple $\pi^0 \gamma$ candidates after all the above criteria, we randomly pick one of them as the best candidate.

To suppress the dominant background from $\psi(3686) \to \pi^+ \pi^- J/\psi$, all possible combinations of $\pi^+ \pi^-$ after the probability-based particle identification are reconstructed. The recoil mass of $\pi^+\pi^-$ is required to be outside of the interval $[m(J/\psi)-7,m(J/\psi)+7]~{\rm MeV}/c^2$. Similarly, to exclude background from $\psi(3686) \to \pi^0 \pi^0 J/\psi$, events are rejected if the recoil mass of any possible $\pi^0 \pi^0$ combination falls in $[m(J/\psi)-15,m(J/\psi)+25]~{\rm MeV}/c^2$. A background contribution may also arise from misidentifying the $\pi^0$ from the decay $\eta \to \pi^+ \pi^- \pi^0$ as the $\pi^0$ from $\psi(3686) \to \pi^0 h_c$, which is suppressed by rejecting events where the invariant mass of any $\pi^+ \pi^- \pi^0$ combination lies within $[m(\eta)-12,m(\eta)+12]~{\rm MeV}/c^2$. The masses of the $J/\psi$ and $\eta$ are taken from the Particle Data Group (PDG)~\cite{ParticleDataGroup:2024cfk}.

\begin{figure}[htbp]
	\centering
	\includegraphics[width=0.4\textwidth]{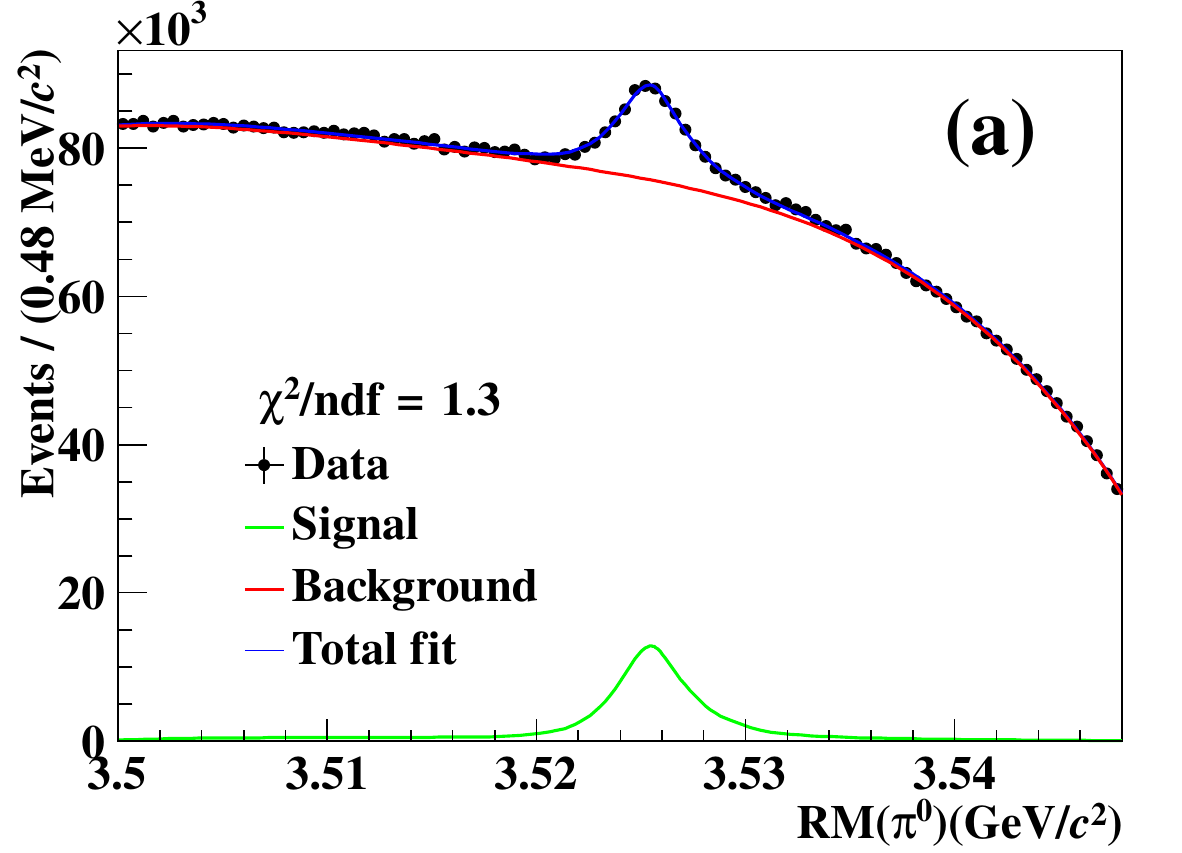}
	\includegraphics[width=0.4\textwidth]{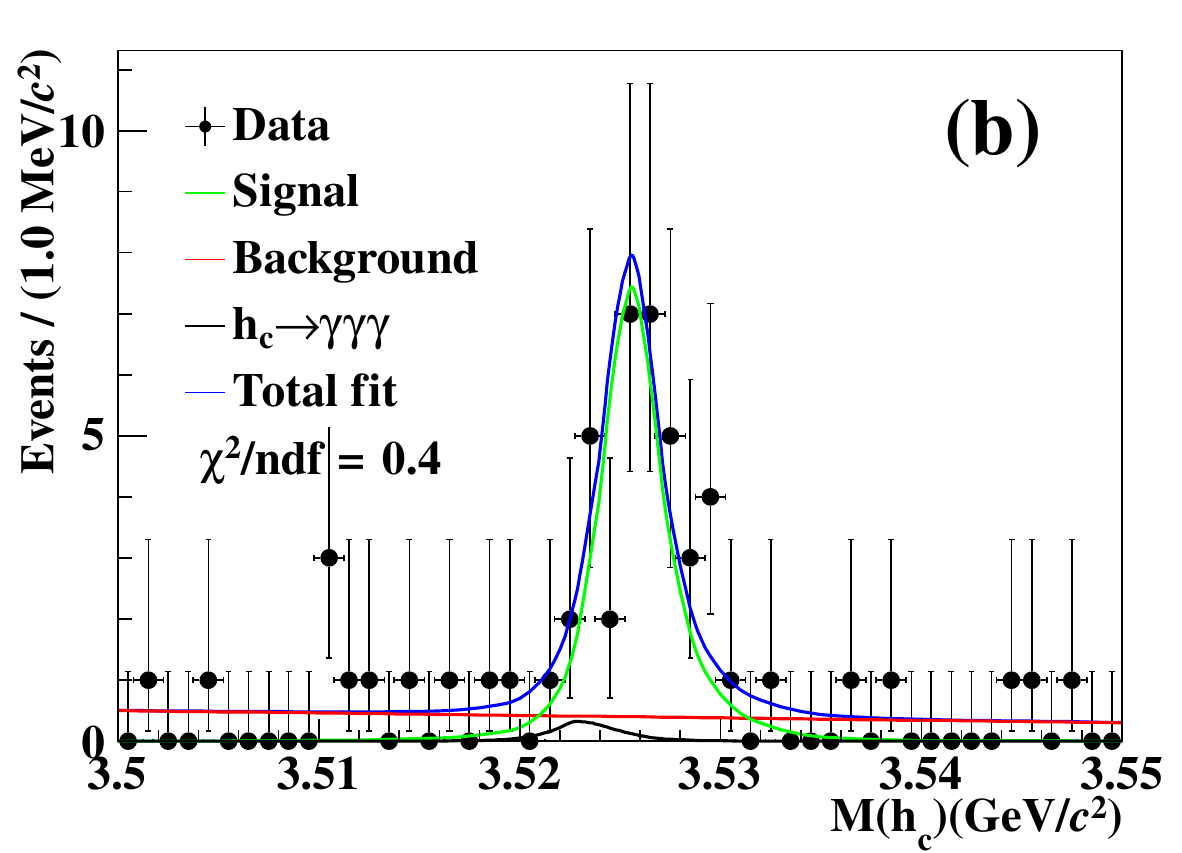}
	\caption{Distributions of (a) $RM(\pi^0)$ from the \textit{tag side} and (b) $M(h_c)$ from the \textit{signal side} with the best fit overlaid. The dots with error bars are data, the blue line is the best fit, the red line is the background, the green line is the signal, and the black line is the peaking background from $h_c \to \gamma\gamma\gamma$.}
	\label{fig:TagSideFit}
\end{figure}

The absolute $\eta_c$ yield on the \textit{tag side} is determined applying a binned extended maximum likelihood fit to the recoil mass spectrum of the selected $\pi^0$ candidates, $RM(\pi^0)$. The fit result is shown in Fig.~\ref{fig:TagSideFit} (a). 
The signal shape is extracted from the signal MC sample and is convoluted with a free Gaussian modeling the discrepancy between simulation and real data, and the background is described by a fourth-order Chebyshev polynomial. The reconstructed signal yield on the \textit{tag side} is determined to be $N_{\rm rec}^{\rm tag} = (158.3 \pm 2.1)\times 10^{3}$. Taking the efficiency $\varepsilon_{\rm rec}^{\rm tag}$ obtained from the signal MC sample as 12.78\%, the inclusive yield of $\eta_c$ from the decay $\psi(3686) \to \pi^0 h_c,~h_c \to \gamma \eta_c$ is calculated to be $N_{\eta_c}^{\rm tag} = (1238.5 \pm 16.6)\times 10^3$.
%%%%%%%%%%%%%%%%%%%%%%%%%%%%%%%% \textit{signal side} %%%%%%%%%%%%%%%%%%%%%%%%%%%%%%%%%%%%%

\begin{figure}[htbp]
	\centering
	\includegraphics[width=0.4\textwidth]{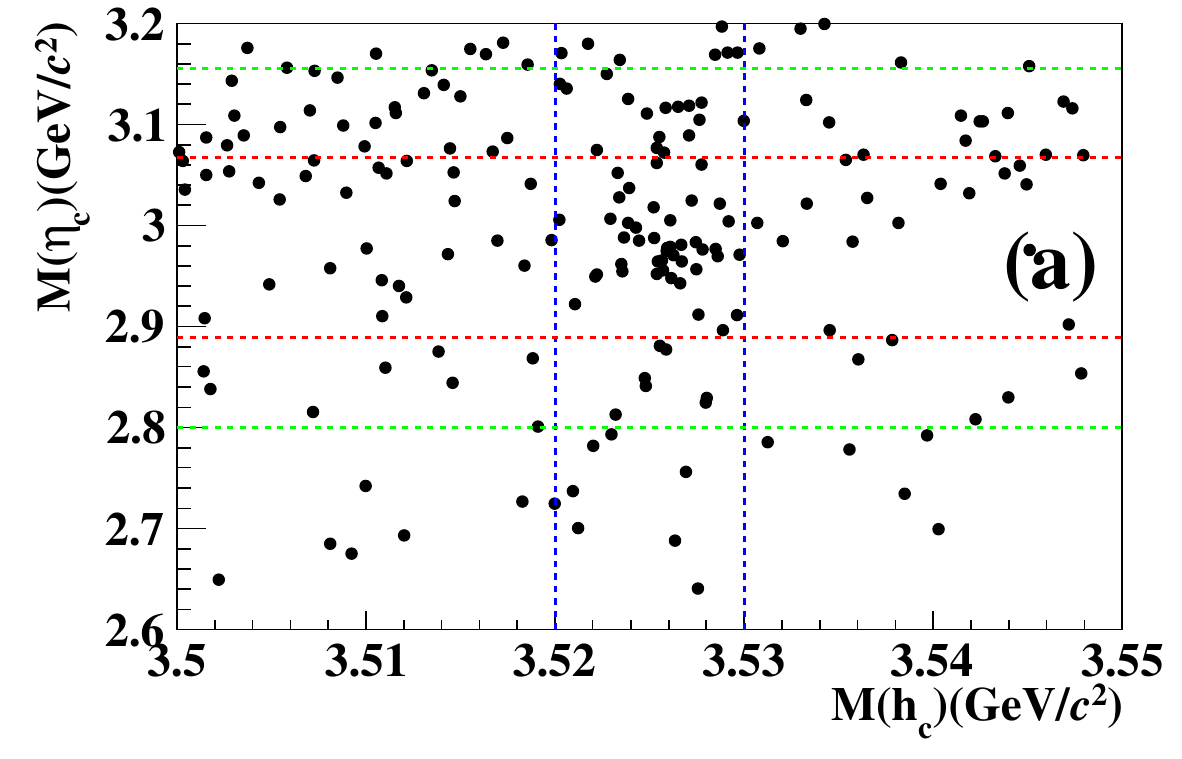}
	\includegraphics[width=0.4\textwidth]{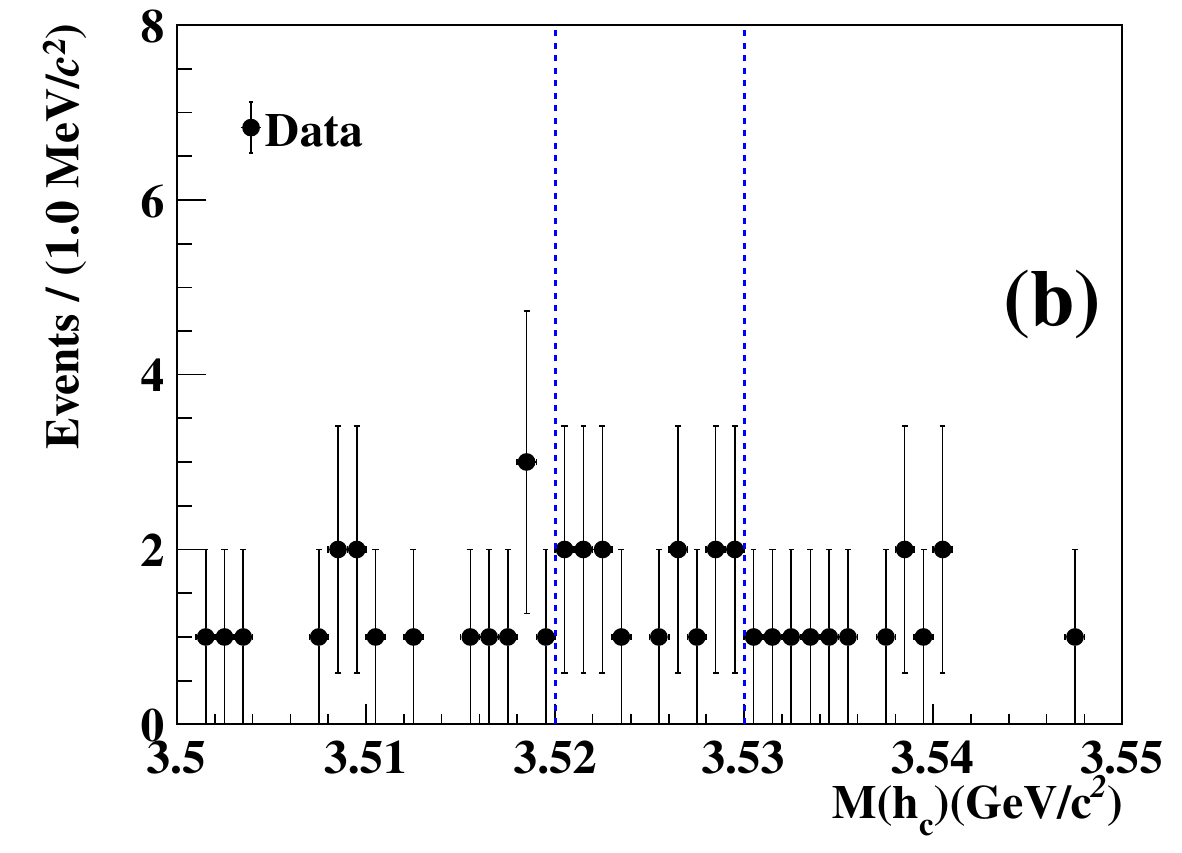}
	\includegraphics[width=0.4\textwidth]{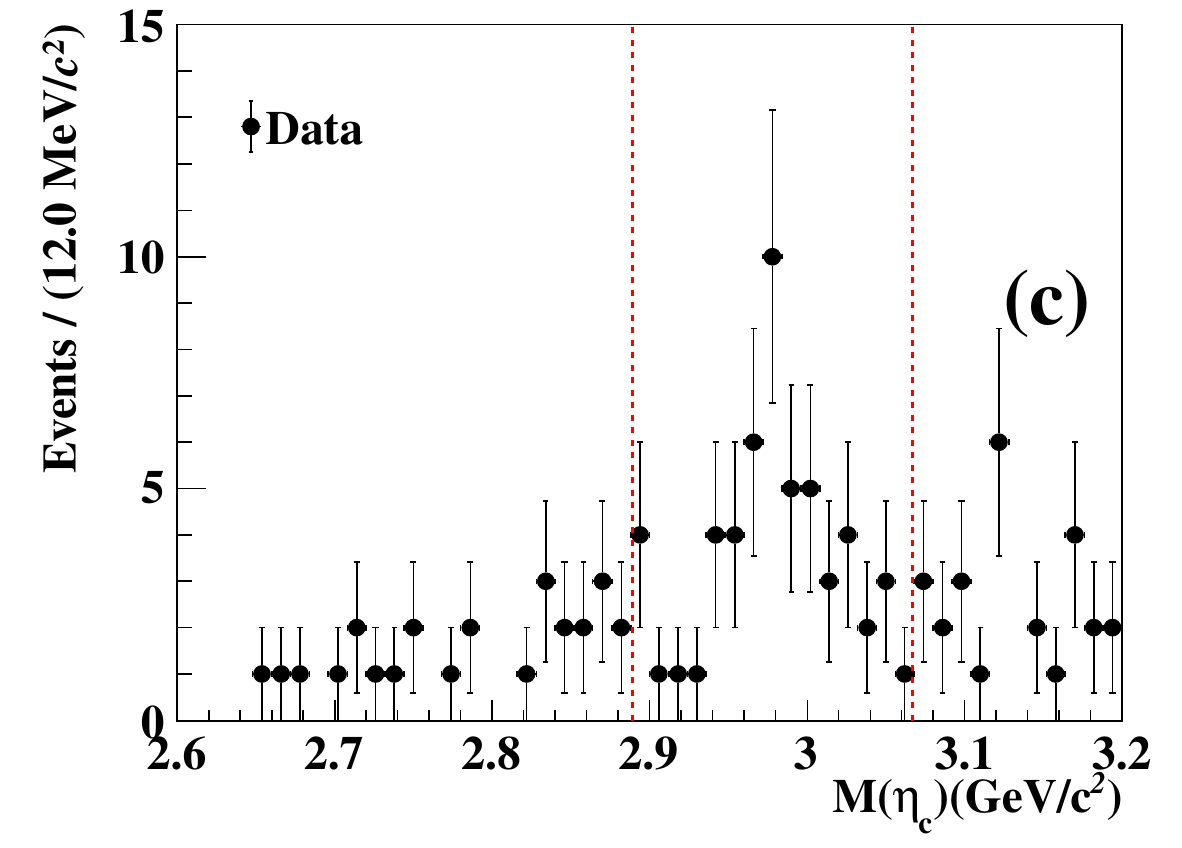}
	\caption{Distributions of (a) $M(h_c)$ versus $M(\eta_c)$, (b) $M(h_c)$ in the $\eta_c$ sideband region and (c) $M(\eta_c)$ in the $h_c$ signal region. The E1 photon selection criteria are not applied in either case. The red and blue dashed lines mark the signal regions for the $\eta_c$ and $h_c$, respectively, while the events outside the green dashed lines are those falling in the $\eta_c$ sideband region.}
	\label{fig:etactohc}
\end{figure}

On the \textit{signal side}, the signal events are reconstructed using the selected events from the \textit{tag side}. The two photons from the decay $\eta_c \to \gamma \gamma$ are picked from the showers that are not used on the \textit{tag side}. A kinematic fit is further performed on all possible $(\pi^0 \gamma)_{\rm tag} \gamma \gamma$ combinations by constraining the total four-momentum (4C) to that of the initial $e^+e^-$ collision. A requirement of $\chi_{\rm 4C}^2 < 70$ is applied to suppress background. If there are multiple candidates left, the best candidate is chosen according to the smallest $\chi^2_{\rm 4C}$. To suppress peaking backgrounds from the same three-photon final states of \(h_c \to \gamma \eta\) and \(h_c \to \gamma \eta'\), we combine the E1 radiative photon with each of the two final-state photons from $\eta_c$ and reject all events where the invariant mass of any photon pair falls within the ranges [0.50, 0.56] GeV/\(c^2\) or [0.88, 0.98] GeV/\(c^2\). Then, the reconstructed signal yield from the \textit{signal side} is extracted from an unbinned extended maximum likelihood fit on the $M(h_c)$ distribution from the final states of $h_c \to \gamma \eta_c,~\eta_c \to \gamma \gamma$ after requiring the invariant mass of the $\gamma \gamma$ from $\eta_c$ decay $M(\eta_c)$ be within $[2889.1, 3066.8]~{\rm MeV}/c^2$, as depicted in Fig.~\ref{fig:TagSideFit} (b).  As shown in Fig.~\ref{fig:etactohc} (c), the signal-to-noise ratio is relatively high and the background shape is complex in the invariant mass spectrum of the $\eta_c$.  We therefore choose to apply the fit to the invariant mass spectrum of the $h_c$. The signal shape is obtained from the signal MC sample, and the background is described by a first-order Chebyshev polynomial. According to Fig.~\ref{fig:etactohc}, a peaking background originating from $h_{c} \to \gamma \gamma \gamma$ is observed in the $h_c$ invariant mass spectrum, but it is not significant. In order to estimate the background of three-photon decays of the $h_c$, the branching fraction of $h_{c} \to \gamma \gamma \gamma$ is estimated using the same $\psi(3686)$ data sample. The candidates for $\psip \to \pi^0 h_c,h_c\to\gamma\gamma\gamma$ are reconstructed by combining the $\pi^0$ candidates using the same selection criteria as those used on the \textit{tag side} with all possible combinations of three photons. A kinematic fit is implemented to constrain the four-momentum of the final-state particles to that of the initial $e^+e^-$ system, and the candidate with the smallest $\chi^2$ is retained. Furthermore, backgrounds from the two-body decay $h_c \to \gamma \eta$, $h_c \to \gamma \eta^{\prime}$ and $h_c \to \gamma \eta_{c}$ are rejected by requiring that no combination of any two photons has an invariant mass within [0.5, 0.6] GeV/$c^2$, [0.9, 1.0] GeV/$c^2$, or above 2.8 GeV/$c^2$. The branching fraction is calculated to be $\mathcal{B}(h_c \to \gamma \gamma \gamma) \times \mathcal{B}(\psip \to \pi^0 h_c) = (3.5 \pm 1.7)\times 10^{-8}$. Thus, $\mathcal{B}(h_c \to \gamma \gamma \gamma) = (4.8 \pm 2.3) \times 10^{-5}$ is calculated by taking $\mathcal{B}(\psip \to \pi^0 h_c)$ from the PDG~\cite{ParticleDataGroup:2024cfk}, where only the statistical uncertainty is considered. The yield from the three-body background is calculated to be $1.4\pm0.7$ using the efficiency obtained from the exclusive MC sample, where the central value is taken as a fixed component in the fit. 
The shape of the background is extracted from an MC sample of $h_{c} \to \gamma \gamma \gamma$. The reconstructed signal yield on the \textit{signal side} is determined to be 
\begin{equation*}
N_{\rm rec}^{\rm sig} = 32.5\pm 6.4.
\end{equation*} 
Taking the efficiency $\varepsilon_{\rm rec}^{\rm sig} = 10.68\%$ obtained from the signal MC sample, the number of produced $\psi(3686) \to \pi^0 h_c,~h_c \to \gamma \eta_c,~\eta_c \to \gamma \gamma$ events is calculated to be $N_{\eta_c}^{\rm sig} = 303.8 \pm 59.9$. Subsequently, the branching fraction is calculated to be 
\begin{equation*}
	\mathcal{B}(\eta_c \to \gamma \gamma) = (2.45 \pm 0.48 \pm 0.09) \times 10^{-4} 
\end{equation*} 
using Eq.~\ref{equ:branching_fraction}, where the first and second uncertainties are statistical and systematic, respectively.

The sources of systematic uncertainty include $\eta_c$ yields and the efficiency of the \textit{tag side}, as well as signal extraction and efficiency of the \textit{signal side}. Since the events from the \textit{signal side} are selected from the events of the \textit{tag side}, the systematic uncertainties on the selection of final particles from the \textit{tag side} are basically cancelled in obtaining $\mathcal{B}(\eta_c \to \gamma \gamma)$. The systematic uncertainty from the reconstruction of the two photons from the $\eta_c$ on the \textit{signal side} remains relevant.  It has been studied thoroughly in a previous BESIII analysis~\cite{BESIII:2024hdv} and evaluated to be no larger than 1.0\% per photon; hence this item of systematic uncertainty is assigned to be 2.0\%. In the selection of the \textit{tag side}, the energy of the E1 photon is constrained. The systematic uncertainty associated with this requirement is tested by varying the criterion around the baseline settings to obtain an alternative measured branching fraction. In the estimation of systematic uncertainties due to the E1 photon, the differences between the results were found to be statistically insignificant compared to their uncertainties, following the Barlow test criterion~\cite{Barlow:2002yb}, so no systematic uncertainty is assigned. Similarly, the systematic uncertainties caused by the exclusion of the backgrounds from $\psi(3686) \to \pi^0 \pi^0 J/\psi$ and $\eta \to \pi^+ \pi^- \pi^0$ are tested and both are negligible. The systematic uncertainty from $\psi(3686) \to \pi^+ \pi^- J/\psi$ has been studied in detail in a previous BESIII analysis~\cite{BESIII:2022jde}, and is determined to be 0.4\%.

The systematic uncertainty caused by the background shape used in the fit of the \textit{tag side} is studied by changing the order of the Chebyshev polynomial in the nominal fit from four to five. The difference from the nominal result is taken as the systematic uncertainty, which is estimated to be 0.3\%. The reconstructed signal yield on the \textit{tag side} is determined by fitting the recoil mass spectrum of the $\pi^0$ ($RM(\pi^0)$), since it has a much better resolution than the recoil mass spectrum of $\pi^0 \gamma$ ($RM(\pi^0\gamma)$). However, the background description is complicated by other decays of the $h_c$ besides $h_c \to \gamma \eta_c$. An alternative way to determine the $\eta_c$ yield from the \textit{tag side} is implemented by fitting the $E_1^{\gamma}$ spectrum, in which the background model becomes different from that in the nominal fit.  This issue is treated as an item of systematic uncertainty, which is estimated as the difference between the nominal fit and the fit to $RM(\pi^0 \gamma)$ and is finally determined to be 1.9\%.

The systematic uncertainty associated with the \textit{signal side} efficiency from the requirement of $\chi^{2}_{\rm 4C}<70$ are found to be negligible. To evaluate the systematic uncertainty arising from the requirement on $M(\eta_c)$, we compare the efficiency-corrected signal yields under different boundary conditions of the region of interest and the largest change from the nominal result is assigned as 1.9\%.

The systematic uncertainty in the signal extraction arising due to the background shape used in the fit is evaluated by replacing the first-order Chebyshev polynomial with an ARGUS function. The resulting deviation from the nominal value is determined as 0.3\%. Additionally, the systematic uncertainty associated with the signal shape in the fit is assessed by comparing the cases with and without the convolution of a Gaussian distribution with free parameters. This contribution is found to be negligible. Furthermore, the systematic uncertainty stemming from the number of peaking background events of $h_{c} \to \gamma \gamma \gamma$ is estimated by varying $N(h_{c} \to \gamma \gamma \gamma)$ within $1\sigma$, namely $N(h_{c} \to \gamma \gamma \gamma) \in [0.7, 2.1]$, and fitting the $M(h_{c})$ distribution again. The maximum deviation from the nominal result is taken as the systematic uncertainty, which is 1.6\%. The potential peaking backgrounds from $h_c \to \gamma\eta(')$ ($\eta(')\to\gamma\gamma$) are evaluated using the known branching fractions from the PDG~\cite{ParticleDataGroup:2024cfk} and the detection efficiencies from the exclusive background MC sample. The number of remaining background events after all selections is found to be negligible and the resulting systematic uncertainty on the branching fraction measurement is 0.9\%. All systematic uncertainties from the three-photon final states are 1.8\%.

All systematic uncertainties considered in our analysis are listed in Table~\ref{tab:sys_summary}. By adding all systematic uncertainties in quadrature, the total systematic uncertainty of the measured branching fraction of $\eta_c \to \gamma \gamma$ is computed to be 3.8\%.
\begin{table}[htbp]
	\centering
	%\doublespacing
	\caption{Summary of relative systematic uncertainties. The symbol \enquote{$-$} denotes an uncertainty value which is negligible}
	\begin{tabular}{lc}
		\hline
		\hline
		Source																		&	Value		\\
		\hline
		Selection of $\pi^{0}$, $\gamma$											&	2.0\%		\\
		E1 $\gamma$ energy requirement		 										&	$-$			\\
		Veto selection 																&	0.4\%		\\
		Background shape in \textit{tag side}										&	0.3\%		\\
		Potential peaking background in \textit{tag side}					&	1.9\%		\\
		$\eta_c$ mass window														&	1.9\%		\\
		$\chi^{2}_{\rm 4C}$ requirement												&	$-$			\\
		Background shape in \textit{signal side}									&	0.3\%		\\
		Signal shape in \textit{signal side} 										&    $-$ 		\\
		Three photon final state background									&	1.8\%		\\
		\hline
		Total																		&	3.8\%		\\
		\hline
		\hline
	\end{tabular}
	\label{tab:sys_summary}
\end{table}

\begin{figure}[htbp]
	\centering
	\includegraphics[width=0.4\textwidth]{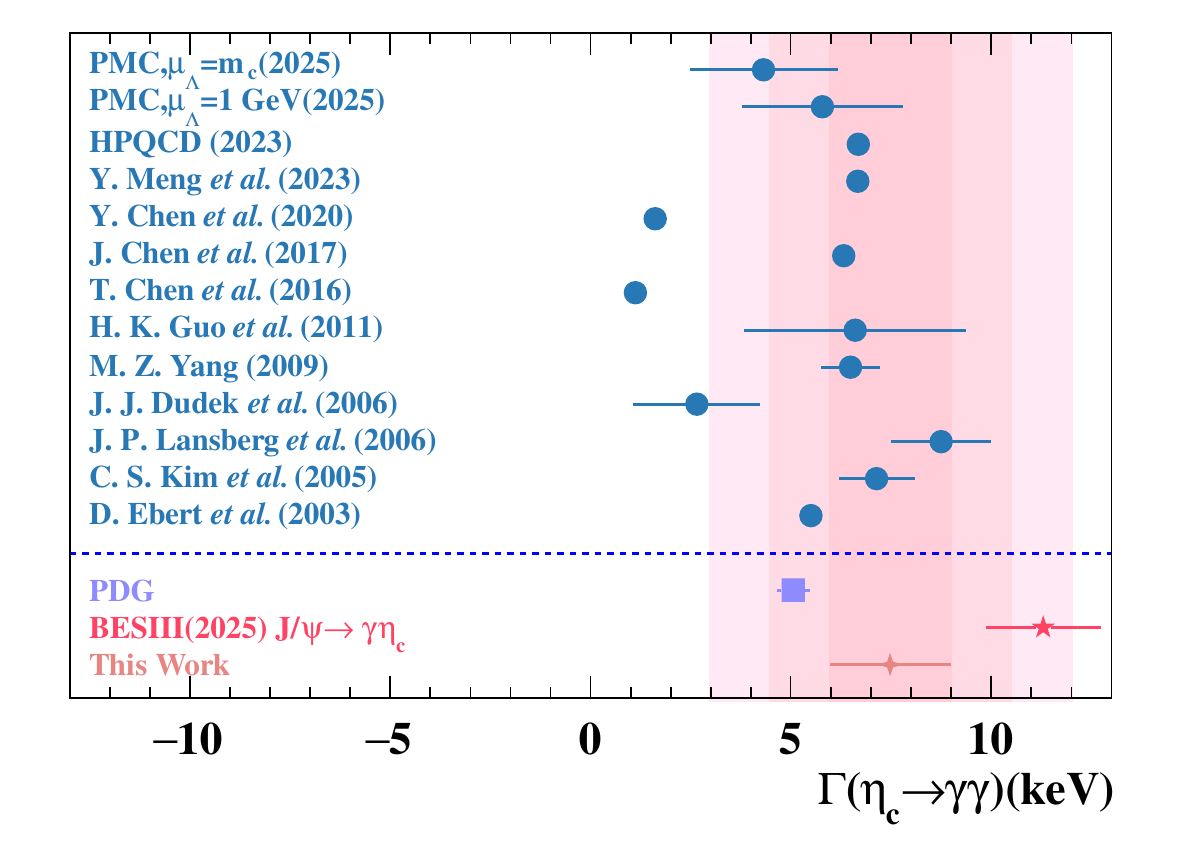}
	\caption{Comparison of experimental measurements and theoretical predictions~\cite{Ebert:2003mu, Kim:2004rz, Lansberg:2006dw, Dudek:2006ut, PhysRevD.79.074026, Guo:2011tz, CLQCD:2016ugl, Chen:2016bpj, Chen_2020,Meng:2021ecs, Colquhoun:2023zbc,Wang_2025}. The coral pink four-pointed star with error bar is the partial width of $\eta_c \to \gamma \gamma$ measured in this analysis, with the pink bands from light to dark showing the ranges from $1\sigma$ to $3\sigma$ of the uncertainty of the measured partial width. The red star with error bar is the partial width reported in another BESIII analysis using the decay $J/\psi \to \gamma \eta_c$~\cite{BESIII:2024rex}. The purple square with error bar is the PDG fitted value with no direct measurements as inputs~\cite{ParticleDataGroup:2024cfk}. 
	The blue dots with error bars are the predictions from theoretical calculations.}
	\label{fig:comparison}
\end{figure}

In summary, we report the first measurement of the absolute branching fraction of the electromagnetic decay $\eta_c \to \gamma \gamma$.  This is based on a sample of about 0.16 million $\eta_c$ events tagged via the process $\psi(3686)\to\pi^0 h_c$, $h_c \to \gamma\eta_c$ using $(2712.4\pm 14.3) \times 10^6~\psi(3686)$ events recorded with the BESIII detector. The model-independent absolute branching fraction measurement is $\mathcal{B}(\eta_c \to \gamma \gamma) = (2.45 \pm 0.48_{\rm stat.} \pm 0.09_{\rm syst.}) \times 10^{-4}$. Taking the total width of the $\eta_c$ to be $\Gamma_{\eta_c} = (30.5\pm 0.5)~{\rm MeV}$ from the PDG~\cite{ParticleDataGroup:2024cfk}, the partial width of $\eta_c \to \gamma \gamma$ is calculated to be $\Gamma(\eta_c \to \gamma \gamma) = (7.48 \pm 1.48_{\rm stat.} \pm 0.30_{\rm syst.})~{\rm keV}$. Our result is consistent with most theoretical estimates and agrees with the most precise calculation from LQCD~\cite{Colquhoun:2023zbc} ($\Gamma(\eta_c \to \gamma \gamma) = (6.788\pm 0.045_{\rm fit}\pm 0.041_{\rm syst.})~{\rm keV}$) and the world average from the PDG~\cite{ParticleDataGroup:2024cfk} ($\Gamma(\eta_c \to \gamma \gamma) = (5.1 \pm 0.4)~{\rm keV}$) within $1.0\sigma$ and $1.6\sigma$, respectively. The various theoretical predictions are shown in Fig~\ref{fig:comparison}. 
Although our result exhibits a $2.5\sigma$ deviation from the product branching fraction, $\mathcal{B}(J/\psi \to \gamma \eta_c) \times \mathcal{B}(\eta_c \to \gamma \gamma)$~\cite{BESIII:2024rex}, reported in 2025 using the $J/\psi \to \gamma \eta_c$ production mechanism, this Letter establishes the first model-independent measurement of $\mathcal{B}(\eta_c \to \gamma \gamma)$. Our determination relies on no external branching fraction inputs, and potential interference effects are negligible due to the large branching fraction of $h_c \to \gamma \eta_c$ ($60 \pm 4\%$)~\cite{ParticleDataGroup:2024cfk}. This result further verifies the QCD calculations, solving the long-standing discrepancy between the theoretical predictions and the direct measurements from BESIII that all rely on $\eta_c$ produced in radiative decays of $J/\psi$. This may imply that potential interference effects in analyses employing radiative decays require more careful and proper handling, and that the branching fraction of $J/\psi \to \gamma \eta_c$ demands more precise measurement.

Taking the leptonic width of the $J/\psi$ to be $\Gamma(J/\psi \to e^+ e^-) = (5.53 \pm 0.10)~{\rm keV}$ from the PDG~\cite{ParticleDataGroup:2024cfk}, the ratio of $\Gamma(J/\psi \to e^+ e^-)$ over $\Gamma(\eta_c \to \gamma \gamma)$ is calculated to be $0.74 \pm 0.15$, which is consistent with the theoretical prediction at the leading order~\cite{Czarnecki:2001zc}. Our experimental result serves as a test for the theoretical models and calculational techniques of charmonium-related physics.

%% Saved at => 2025-08-28
\textbf{Acknowledgement}

The BESIII Collaboration thanks the staff of BEPCII (https://cstr.cn/31109.02.BEPC) and the IHEP computing center for their strong support. This work is supported in part by National Key R\&D Program of China under Contracts Nos. 2025YFA1613900, 2023YFA1606000, 2023YFA1606704; National Natural Science Foundation of China (NSFC) under Contracts Nos. 11635010, 11935015, 11935016, 11935018, 12025502, 12035009, 12035013, 12061131003, 12192260, 12192261, 12192262, 12192263, 12192264, 12192265, 12221005, 12225509, 12235017, 12361141819; the Chinese Academy of Sciences (CAS) Large-Scale Scientific Facility Program; CAS Project for Young Scientists in Basic Re-search No. YSBR-117; the Strategic Priority Research Program of Chinese Academy of Sciences under Contract No. XDA0480600; CAS under Contract No. YSBR-101; 100 Talents Program of CAS; The Institute of Nuclear and Particle Physics (INPAC) and Shanghai Key Laboratory for Particle Physics and Cosmology; ERC under Contract No. 758462; German Research Foundation DFG under Contract No. FOR5327; Istituto Nazionale di Fisica Nucleare, Italy; Knut and Alice Wallenberg Foundation under Contracts Nos. 2021.0174, 2021.0299, 2023.0315; Ministry of Development of Turkey under Contract No. DPT2006K-120470; National Research Foundation of Korea under Contract No. NRF-2022R1A2C1092335; National Science and Technology fund of Mongolia; Polish National Science Centre under Contract No. 2024/53/B/ST2/00975; STFC (United Kingdom); Swedish Research Council under Contract No. 2019.04595; U. S. Department of Energy under Contract No. DE-FG02-05ER41374

%\textbf{Other Fund Information}

%To be inserted with an additional sentence into papers that are relevant to the topic of special funding for specific topics. Authors can suggest which to Li Weiguo and/or the physics coordinator.
        %Example added sentence: This paper is also supported by the NSFC under Contract Nos. 10805053, 10979059, ....National Natural Science Foundation of China (NSFC), 10805053, PWANational Natural Science Foundation of China (NSFC), 10979059, Lund弦碎裂强子化模型及其通用强子产生器研究National Natural Science Foundation of China (NSFC), 10775075, National Natural Science Foundation of China (NSFC), 10979012, baryonsNational Natural Science Foundation of China (NSFC), 10979038, charmoniumNational Natural Science Foundation of China (NSFC), 10905034, psi(2S)->B BbarNational Natural Science Foundation of China (NSFC), 10975093, D 介子National Natural Science Foundation of China (NSFC), 10979033, psi(2S)->VPNational Natural Science Foundation of China (NSFC), 10979058, hcNational Natural Science Foundation of China (NSFC), 10975143, charmonium rare decays
%% ends here %%

\bibliography{references}

%apsrev4-2.bst 2019-01-14 (MD) hand-edited version of apsrev4-1.bst
%Control: key (0)
%Control: author (8) initials jnrlst
%Control: editor formatted (1) identically to author
%Control: production of article title (0) allowed
%Control: page (0) single
%Control: year (1) truncated
%Control: production of eprint (0) enabled
\begin{thebibliography}{51}%
\makeatletter
\providecommand \@ifxundefined [1]{%
 \@ifx{#1\undefined}
}%
\providecommand \@ifnum [1]{%
 \ifnum #1\expandafter \@firstoftwo
 \else \expandafter \@secondoftwo
 \fi
}%
\providecommand \@ifx [1]{%
 \ifx #1\expandafter \@firstoftwo
 \else \expandafter \@secondoftwo
 \fi
}%
\providecommand \natexlab [1]{#1}%
\providecommand \enquote  [1]{``#1''}%
\providecommand \bibnamefont  [1]{#1}%
\providecommand \bibfnamefont [1]{#1}%
\providecommand \citenamefont [1]{#1}%
\providecommand \href@noop [0]{\@secondoftwo}%
\providecommand \href [0]{\begingroup \@sanitize@url \@href}%
\providecommand \@href[1]{\@@startlink{#1}\@@href}%
\providecommand \@@href[1]{\endgroup#1\@@endlink}%
\providecommand \@sanitize@url [0]{\catcode `\\12\catcode `\$12\catcode
  `\&12\catcode `\#12\catcode `\^12\catcode `\_12\catcode `\%12\relax}%
\providecommand \@@startlink[1]{}%
\providecommand \@@endlink[0]{}%
\providecommand \url  [0]{\begingroup\@sanitize@url \@url }%
\providecommand \@url [1]{\endgroup\@href {#1}{\urlprefix }}%
\providecommand \urlprefix  [0]{URL }%
\providecommand \Eprint [0]{\href }%
\providecommand \doibase [0]{https://doi.org/}%
\providecommand \selectlanguage [0]{\@gobble}%
\providecommand \bibinfo  [0]{\@secondoftwo}%
\providecommand \bibfield  [0]{\@secondoftwo}%
\providecommand \translation [1]{[#1]}%
\providecommand \BibitemOpen [0]{}%
\providecommand \bibitemStop [0]{}%
\providecommand \bibitemNoStop [0]{.\EOS\space}%
\providecommand \EOS [0]{\spacefactor3000\relax}%
\providecommand \BibitemShut  [1]{\csname bibitem#1\endcsname}%
\let\auto@bib@innerbib\@empty
%</preamble>
\bibitem [{\citenamefont {Czarnecki}\ and\ \citenamefont
  {Melnikov}(2001)}]{Czarnecki:2001zc}%
  \BibitemOpen
  \bibfield  {author} {\bibinfo {author} {\bibfnamefont {A.}~\bibnamefont
  {Czarnecki}}\ and\ \bibinfo {author} {\bibfnamefont {K.}~\bibnamefont
  {Melnikov}},\ }\href {https://doi.org/10.1016/S0370-2693(01)01129-7}
  {\bibfield  {journal} {\bibinfo  {journal} {Phys. Lett. B}\ }\textbf
  {\bibinfo {volume} {519}},\ \bibinfo {pages} {212} (\bibinfo {year}
  {2001})}\BibitemShut {NoStop}%
\bibitem [{\citenamefont {Bodwin}\ and\ \citenamefont
  {Petrelli}(2002)}]{Bodwin:2002cfe}%
  \BibitemOpen
  \bibfield  {author} {\bibinfo {author} {\bibfnamefont {G.~T.}\ \bibnamefont
  {Bodwin}}\ and\ \bibinfo {author} {\bibfnamefont {A.}~\bibnamefont
  {Petrelli}},\ }\href {https://doi.org/10.1103/PhysRevD.66.094011} {\bibfield
  {journal} {\bibinfo  {journal} {Phys. Rev. D}\ }\textbf {\bibinfo {volume}
  {66}},\ \bibinfo {pages} {094011} (\bibinfo {year} {2002})},\ \bibinfo {note}
  {[Erratum: Phys.Rev.D 87, 039902 (2013)]}\BibitemShut {NoStop}%
\bibitem [{\citenamefont {Brambilla}\ \emph {et~al.}(2006)\citenamefont
  {Brambilla}, \citenamefont {Mereghetti},\ and\ \citenamefont
  {Vairo}}]{Brambilla:2006ph}%
  \BibitemOpen
  \bibfield  {author} {\bibinfo {author} {\bibfnamefont {N.}~\bibnamefont
  {Brambilla}}, \bibinfo {author} {\bibfnamefont {E.}~\bibnamefont
  {Mereghetti}},\ and\ \bibinfo {author} {\bibfnamefont {A.}~\bibnamefont
  {Vairo}},\ }\href {https://doi.org/10.1088/1126-6708/2006/08/039} {\bibfield
  {journal} {\bibinfo  {journal} {J. High Energy Phys.}\ }\textbf {\bibinfo
  {volume} {08}},\ \bibinfo {pages} {039}},\ \bibinfo {note} {[Erratum: JHEP
  04, 058 (2011)]}\BibitemShut {NoStop}%
\bibitem [{\citenamefont {Guo}\ \emph {et~al.}(2011)\citenamefont {Guo},
  \citenamefont {Ma},\ and\ \citenamefont {Chao}}]{Guo:2011tz}%
  \BibitemOpen
  \bibfield  {author} {\bibinfo {author} {\bibfnamefont {H.-K.}\ \bibnamefont
  {Guo}}, \bibinfo {author} {\bibfnamefont {Y.-Q.}\ \bibnamefont {Ma}},\ and\
  \bibinfo {author} {\bibfnamefont {K.-T.}\ \bibnamefont {Chao}},\ }\href
  {https://doi.org/10.1103/PhysRevD.83.114038} {\bibfield  {journal} {\bibinfo
  {journal} {Phys. Rev. D}\ }\textbf {\bibinfo {volume} {83}},\ \bibinfo
  {pages} {114038} (\bibinfo {year} {2011})}\BibitemShut {NoStop}%
\bibitem [{\citenamefont {Jia}\ \emph {et~al.}(2011)\citenamefont {Jia},
  \citenamefont {Yang}, \citenamefont {Sang},\ and\ \citenamefont
  {Xu}}]{Jia:2011ah}%
  \BibitemOpen
  \bibfield  {author} {\bibinfo {author} {\bibfnamefont {Y.}~\bibnamefont
  {Jia}}, \bibinfo {author} {\bibfnamefont {X.-T.}\ \bibnamefont {Yang}},
  \bibinfo {author} {\bibfnamefont {W.-L.}\ \bibnamefont {Sang}},\ and\
  \bibinfo {author} {\bibfnamefont {J.}~\bibnamefont {Xu}},\ }\href
  {https://doi.org/10.1007/JHEP06(2011)097} {\bibfield  {journal} {\bibinfo
  {journal} {JHEP}\ }\textbf {\bibinfo {volume} {06}},\ \bibinfo {pages}
  {097}}\BibitemShut {NoStop}%
\bibitem [{\citenamefont {Feng}\ \emph {et~al.}(2017)\citenamefont {Feng},
  \citenamefont {Jia},\ and\ \citenamefont {Sang}}]{Feng:2017hlu}%
  \BibitemOpen
  \bibfield  {author} {\bibinfo {author} {\bibfnamefont {F.}~\bibnamefont
  {Feng}}, \bibinfo {author} {\bibfnamefont {Y.}~\bibnamefont {Jia}},\ and\
  \bibinfo {author} {\bibfnamefont {W.-L.}\ \bibnamefont {Sang}},\ }\href
  {https://doi.org/10.1103/PhysRevLett.119.252001} {\bibfield  {journal}
  {\bibinfo  {journal} {Phys. Rev. Lett.}\ }\textbf {\bibinfo {volume} {119}},\
  \bibinfo {pages} {252001} (\bibinfo {year} {2017})}\BibitemShut {NoStop}%
\bibitem [{\citenamefont {Brambilla}\ \emph {et~al.}(2018)\citenamefont
  {Brambilla}, \citenamefont {Chung},\ and\ \citenamefont
  {Komijani}}]{Brambilla:2018tyu}%
  \BibitemOpen
  \bibfield  {author} {\bibinfo {author} {\bibfnamefont {N.}~\bibnamefont
  {Brambilla}}, \bibinfo {author} {\bibfnamefont {H.~S.}\ \bibnamefont
  {Chung}},\ and\ \bibinfo {author} {\bibfnamefont {J.}~\bibnamefont
  {Komijani}},\ }\href {https://doi.org/10.1103/PhysRevD.98.114020} {\bibfield
  {journal} {\bibinfo  {journal} {Phys. Rev. D}\ }\textbf {\bibinfo {volume}
  {98}},\ \bibinfo {pages} {114020} (\bibinfo {year} {2018})}\BibitemShut
  {NoStop}%
\bibitem [{\citenamefont {Yu}\ \emph {et~al.}(2020)\citenamefont {Yu},
  \citenamefont {Wu}, \citenamefont {Zeng}, \citenamefont {Huang},\ and\
  \citenamefont {Yu}}]{Yu:2019mce}%
  \BibitemOpen
  \bibfield  {author} {\bibinfo {author} {\bibfnamefont {Q.}~\bibnamefont
  {Yu}}, \bibinfo {author} {\bibfnamefont {X.-G.}\ \bibnamefont {Wu}}, \bibinfo
  {author} {\bibfnamefont {J.}~\bibnamefont {Zeng}}, \bibinfo {author}
  {\bibfnamefont {X.-D.}\ \bibnamefont {Huang}},\ and\ \bibinfo {author}
  {\bibfnamefont {H.-M.}\ \bibnamefont {Yu}},\ }\href
  {https://doi.org/10.1140/epjc/s10052-020-7967-x} {\bibfield  {journal}
  {\bibinfo  {journal} {Eur. Phys. J. C}\ }\textbf {\bibinfo {volume} {80}},\
  \bibinfo {pages} {362} (\bibinfo {year} {2020})}\BibitemShut {NoStop}%
\bibitem [{\citenamefont {Abreu}\ \emph {et~al.}(2023)\citenamefont {Abreu},
  \citenamefont {Becchetti}, \citenamefont {Duhr},\ and\ \citenamefont
  {Ozcelik}}]{Abreu:2022cco}%
  \BibitemOpen
  \bibfield  {author} {\bibinfo {author} {\bibfnamefont {S.}~\bibnamefont
  {Abreu}}, \bibinfo {author} {\bibfnamefont {M.}~\bibnamefont {Becchetti}},
  \bibinfo {author} {\bibfnamefont {C.}~\bibnamefont {Duhr}},\ and\ \bibinfo
  {author} {\bibfnamefont {M.~A.}\ \bibnamefont {Ozcelik}},\ }\href
  {https://doi.org/10.1007/JHEP02(2023)250} {\bibfield  {journal} {\bibinfo
  {journal} {JHEP}\ }\textbf {\bibinfo {volume} {02}},\ \bibinfo {pages}
  {250}}\BibitemShut {NoStop}%
\bibitem [{\citenamefont {Ackleh}\ and\ \citenamefont
  {Barnes}(1992)}]{Ackleh:1991dy}%
  \BibitemOpen
  \bibfield  {author} {\bibinfo {author} {\bibfnamefont {E.~S.}\ \bibnamefont
  {Ackleh}}\ and\ \bibinfo {author} {\bibfnamefont {T.}~\bibnamefont
  {Barnes}},\ }\href {https://doi.org/10.1103/PhysRevD.45.232} {\bibfield
  {journal} {\bibinfo  {journal} {Phys. Rev. D}\ }\textbf {\bibinfo {volume}
  {45}},\ \bibinfo {pages} {232} (\bibinfo {year} {1992})}\BibitemShut
  {NoStop}%
\bibitem [{\citenamefont {Munz}(1996)}]{Munz:1996hb}%
  \BibitemOpen
  \bibfield  {author} {\bibinfo {author} {\bibfnamefont {C.~R.}\ \bibnamefont
  {Munz}},\ }\href {https://doi.org/10.1016/S0375-9474(96)00265-5} {\bibfield
  {journal} {\bibinfo  {journal} {Nucl. Phys. A}\ }\textbf {\bibinfo {volume}
  {609}},\ \bibinfo {pages} {364} (\bibinfo {year} {1996})}\BibitemShut
  {NoStop}%
\bibitem [{\citenamefont {Ebert}\ \emph {et~al.}(2003)\citenamefont {Ebert},
  \citenamefont {Faustov},\ and\ \citenamefont {Galkin}}]{Ebert:2003mu}%
  \BibitemOpen
  \bibfield  {author} {\bibinfo {author} {\bibfnamefont {D.}~\bibnamefont
  {Ebert}}, \bibinfo {author} {\bibfnamefont {R.~N.}\ \bibnamefont {Faustov}},\
  and\ \bibinfo {author} {\bibfnamefont {V.~O.}\ \bibnamefont {Galkin}},\
  }\href {https://doi.org/10.1142/S021773230300971X} {\bibfield  {journal}
  {\bibinfo  {journal} {Mod. Phys. Lett. A}\ }\textbf {\bibinfo {volume}
  {18}},\ \bibinfo {pages} {601} (\bibinfo {year} {2003})}\BibitemShut
  {NoStop}%
\bibitem [{\citenamefont {Kim}\ \emph {et~al.}(2005)\citenamefont {Kim},
  \citenamefont {Lee},\ and\ \citenamefont {Wang}}]{Kim:2004rz}%
  \BibitemOpen
  \bibfield  {author} {\bibinfo {author} {\bibfnamefont {C.~S.}\ \bibnamefont
  {Kim}}, \bibinfo {author} {\bibfnamefont {T.}~\bibnamefont {Lee}},\ and\
  \bibinfo {author} {\bibfnamefont {G.-L.}\ \bibnamefont {Wang}},\ }\href
  {https://doi.org/10.1016/j.physletb.2004.11.084} {\bibfield  {journal}
  {\bibinfo  {journal} {Phys. Lett. B}\ }\textbf {\bibinfo {volume} {606}},\
  \bibinfo {pages} {323} (\bibinfo {year} {2005})}\BibitemShut {NoStop}%
\bibitem [{\citenamefont {Chao}\ \emph {et~al.}(1997)\citenamefont {Chao},
  \citenamefont {Huang}, \citenamefont {Liu},\ and\ \citenamefont
  {Tang}}]{Chao:1996bj}%
  \BibitemOpen
  \bibfield  {author} {\bibinfo {author} {\bibfnamefont {K.-T.}\ \bibnamefont
  {Chao}}, \bibinfo {author} {\bibfnamefont {H.-W.}\ \bibnamefont {Huang}},
  \bibinfo {author} {\bibfnamefont {J.-H.}\ \bibnamefont {Liu}},\ and\ \bibinfo
  {author} {\bibfnamefont {J.}~\bibnamefont {Tang}},\ }\href
  {https://doi.org/10.1103/PhysRevD.56.368} {\bibfield  {journal} {\bibinfo
  {journal} {Phys. Rev. D}\ }\textbf {\bibinfo {volume} {56}},\ \bibinfo
  {pages} {368} (\bibinfo {year} {1997})}\BibitemShut {NoStop}%
\bibitem [{\citenamefont {Chen}\ \emph {et~al.}(2017)\citenamefont {Chen},
  \citenamefont {Ding}, \citenamefont {Chang},\ and\ \citenamefont
  {Liu}}]{Chen:2016bpj}%
  \BibitemOpen
  \bibfield  {author} {\bibinfo {author} {\bibfnamefont {J.}~\bibnamefont
  {Chen}}, \bibinfo {author} {\bibfnamefont {M.}~\bibnamefont {Ding}}, \bibinfo
  {author} {\bibfnamefont {L.}~\bibnamefont {Chang}},\ and\ \bibinfo {author}
  {\bibfnamefont {Y.-x.}\ \bibnamefont {Liu}},\ }\href
  {https://doi.org/10.1103/PhysRevD.95.016010} {\bibfield  {journal} {\bibinfo
  {journal} {Phys. Rev. D}\ }\textbf {\bibinfo {volume} {95}},\ \bibinfo
  {pages} {016010} (\bibinfo {year} {2017})}\BibitemShut {NoStop}%
\bibitem [{\citenamefont {Higuera-Angulo}\ \emph {et~al.}(2024)\citenamefont
  {Higuera-Angulo}, \citenamefont {Hern\'andez-Pinto}, \citenamefont {Raya},\
  and\ \citenamefont {Bashir}}]{Higuera-Angulo:2024oui}%
  \BibitemOpen
  \bibfield  {author} {\bibinfo {author} {\bibfnamefont {I.~M.}\ \bibnamefont
  {Higuera-Angulo}}, \bibinfo {author} {\bibfnamefont {R.~J.}\ \bibnamefont
  {Hern\'andez-Pinto}}, \bibinfo {author} {\bibfnamefont {K.}~\bibnamefont
  {Raya}},\ and\ \bibinfo {author} {\bibfnamefont {A.}~\bibnamefont {Bashir}},\
  }\href {https://doi.org/10.1103/PhysRevD.110.034013} {\bibfield  {journal}
  {\bibinfo  {journal} {Phys. Rev. D}\ }\textbf {\bibinfo {volume} {110}},\
  \bibinfo {pages} {034013} (\bibinfo {year} {2024})}\BibitemShut {NoStop}%
\bibitem [{\citenamefont {Yang}(2009)}]{PhysRevD.79.074026}%
  \BibitemOpen
  \bibfield  {author} {\bibinfo {author} {\bibfnamefont {M.-Z.}\ \bibnamefont
  {Yang}},\ }\href {https://doi.org/10.1103/PhysRevD.79.074026} {\bibfield
  {journal} {\bibinfo  {journal} {Phys. Rev. D}\ }\textbf {\bibinfo {volume}
  {79}},\ \bibinfo {pages} {074026} (\bibinfo {year} {2009})}\BibitemShut
  {NoStop}%
\bibitem [{\citenamefont {Lansberg}\ and\ \citenamefont
  {Pham}(2006)}]{Lansberg:2006dw}%
  \BibitemOpen
  \bibfield  {author} {\bibinfo {author} {\bibfnamefont {J.~P.}\ \bibnamefont
  {Lansberg}}\ and\ \bibinfo {author} {\bibfnamefont {T.~N.}\ \bibnamefont
  {Pham}},\ }\href {https://doi.org/10.1103/PhysRevD.74.034001} {\bibfield
  {journal} {\bibinfo  {journal} {Phys. Rev. D}\ }\textbf {\bibinfo {volume}
  {74}},\ \bibinfo {pages} {034001} (\bibinfo {year} {2006})}\BibitemShut
  {NoStop}%
\bibitem [{\citenamefont {Li}\ \emph {et~al.}(2020)\citenamefont {Li},
  \citenamefont {Feng},\ and\ \citenamefont {Ma}}]{Li:2019ncs}%
  \BibitemOpen
  \bibfield  {author} {\bibinfo {author} {\bibfnamefont {R.}~\bibnamefont
  {Li}}, \bibinfo {author} {\bibfnamefont {Y.}~\bibnamefont {Feng}},\ and\
  \bibinfo {author} {\bibfnamefont {Y.-Q.}\ \bibnamefont {Ma}},\ }\href
  {https://doi.org/10.1007/JHEP05(2020)009} {\bibfield  {journal} {\bibinfo
  {journal} {JHEP}\ }\textbf {\bibinfo {volume} {05}},\ \bibinfo {pages}
  {009}}\BibitemShut {NoStop}%
\bibitem [{\citenamefont {Dudek}\ and\ \citenamefont
  {Edwards}(2006)}]{Dudek:2006ut}%
  \BibitemOpen
  \bibfield  {author} {\bibinfo {author} {\bibfnamefont {J.~J.}\ \bibnamefont
  {Dudek}}\ and\ \bibinfo {author} {\bibfnamefont {R.~G.}\ \bibnamefont
  {Edwards}},\ }\href {https://doi.org/10.1103/PhysRevLett.97.172001}
  {\bibfield  {journal} {\bibinfo  {journal} {Phys. Rev. Lett.}\ }\textbf
  {\bibinfo {volume} {97}},\ \bibinfo {pages} {172001} (\bibinfo {year}
  {2006})}\BibitemShut {NoStop}%
\bibitem [{\citenamefont {Chen}\ \emph {et~al.}(2016)\citenamefont {Chen} \emph
  {et~al.}}]{CLQCD:2016ugl}%
  \BibitemOpen
  \bibfield  {author} {\bibinfo {author} {\bibfnamefont {T.}~\bibnamefont
  {Chen}} \emph {et~al.} (\bibinfo {collaboration} {CLQCD Collaboration}),\
  }\href {https://doi.org/10.1140/epjc/s10052-016-4212-8} {\bibfield  {journal}
  {\bibinfo  {journal} {Eur. Phys. J. C}\ }\textbf {\bibinfo {volume} {76}},\
  \bibinfo {pages} {358} (\bibinfo {year} {2016})}\BibitemShut {NoStop}%
\bibitem [{\citenamefont {Chen}\ \emph
  {et~al.}(2020{\natexlab{a}})\citenamefont {Chen}, \citenamefont {Gong},
  \citenamefont {Li}, \citenamefont {Liu}, \citenamefont {Liu}, \citenamefont
  {Liu}, \citenamefont {Ma}, \citenamefont {Meng}, \citenamefont {Xiong},\ and\
  \citenamefont {Zhang}}]{CLQCD:2020njc}%
  \BibitemOpen
  \bibfield  {author} {\bibinfo {author} {\bibfnamefont {Y.}~\bibnamefont
  {Chen}}, \bibinfo {author} {\bibfnamefont {M.}~\bibnamefont {Gong}}, \bibinfo
  {author} {\bibfnamefont {N.}~\bibnamefont {Li}}, \bibinfo {author}
  {\bibfnamefont {C.}~\bibnamefont {Liu}}, \bibinfo {author} {\bibfnamefont
  {Y.-B.}\ \bibnamefont {Liu}}, \bibinfo {author} {\bibfnamefont
  {Z.}~\bibnamefont {Liu}}, \bibinfo {author} {\bibfnamefont {J.-P.}\
  \bibnamefont {Ma}}, \bibinfo {author} {\bibfnamefont {Y.}~\bibnamefont
  {Meng}}, \bibinfo {author} {\bibfnamefont {C.}~\bibnamefont {Xiong}},\ and\
  \bibinfo {author} {\bibfnamefont {K.-L.}\ \bibnamefont {Zhang}} (\bibinfo
  {collaboration} {CLQCD Collaboration}),\ }\href
  {https://doi.org/10.1088/1674-1137/44/8/083108} {\bibfield  {journal}
  {\bibinfo  {journal} {Chin. Phys. C}\ }\textbf {\bibinfo {volume} {44}},\
  \bibinfo {pages} {083108} (\bibinfo {year} {2020}{\natexlab{a}})},\ \bibinfo
  {note} {[Erratum: Chin.Phys.C 46, 059001 (2022)]}\BibitemShut {NoStop}%
\bibitem [{\citenamefont {Liu}\ \emph {et~al.}(2020)\citenamefont {Liu},
  \citenamefont {Meng},\ and\ \citenamefont {Zhang}}]{Liu:2020qfz}%
  \BibitemOpen
  \bibfield  {author} {\bibinfo {author} {\bibfnamefont {C.}~\bibnamefont
  {Liu}}, \bibinfo {author} {\bibfnamefont {Y.}~\bibnamefont {Meng}},\ and\
  \bibinfo {author} {\bibfnamefont {K.-L.}\ \bibnamefont {Zhang}},\ }\href
  {https://doi.org/10.1103/PhysRevD.102.034502} {\bibfield  {journal} {\bibinfo
   {journal} {Phys. Rev. D}\ }\textbf {\bibinfo {volume} {102}},\ \bibinfo
  {pages} {034502} (\bibinfo {year} {2020})}\BibitemShut {NoStop}%
\bibitem [{\citenamefont {Zhang}\ \emph {et~al.}(2022)\citenamefont {Zhang},
  \citenamefont {Sun}, \citenamefont {Chen}, \citenamefont {Gong},
  \citenamefont {Gui},\ and\ \citenamefont {Liu}}]{Zhang:2021xvl}%
  \BibitemOpen
  \bibfield  {author} {\bibinfo {author} {\bibfnamefont {R.}~\bibnamefont
  {Zhang}}, \bibinfo {author} {\bibfnamefont {W.}~\bibnamefont {Sun}}, \bibinfo
  {author} {\bibfnamefont {Y.}~\bibnamefont {Chen}}, \bibinfo {author}
  {\bibfnamefont {M.}~\bibnamefont {Gong}}, \bibinfo {author} {\bibfnamefont
  {L.-C.}\ \bibnamefont {Gui}},\ and\ \bibinfo {author} {\bibfnamefont
  {Z.}~\bibnamefont {Liu}},\ }\href
  {https://doi.org/10.1016/j.physletb.2022.136960} {\bibfield  {journal}
  {\bibinfo  {journal} {Phys. Lett. B}\ }\textbf {\bibinfo {volume} {827}},\
  \bibinfo {pages} {136960} (\bibinfo {year} {2022})}\BibitemShut {NoStop}%
\bibitem [{\citenamefont {Colquhoun}\ \emph {et~al.}(2023)\citenamefont
  {Colquhoun}, \citenamefont {Cooper}, \citenamefont {Davies},\ and\
  \citenamefont {Lepage}}]{Colquhoun:2023zbc}%
  \BibitemOpen
  \bibfield  {author} {\bibinfo {author} {\bibfnamefont {B.}~\bibnamefont
  {Colquhoun}}, \bibinfo {author} {\bibfnamefont {L.~J.}\ \bibnamefont
  {Cooper}}, \bibinfo {author} {\bibfnamefont {C.~T.~H.}\ \bibnamefont
  {Davies}},\ and\ \bibinfo {author} {\bibfnamefont {G.~P.}\ \bibnamefont
  {Lepage}} (\bibinfo {collaboration} {HPQCD Collaboration}),\ }\href
  {https://doi.org/10.1103/PhysRevD.108.014513} {\bibfield  {journal} {\bibinfo
   {journal} {Phys. Rev. D}\ }\textbf {\bibinfo {volume} {108}},\ \bibinfo
  {pages} {014513} (\bibinfo {year} {2023})}\BibitemShut {NoStop}%
\bibitem [{\citenamefont {Meng}\ \emph {et~al.}(2023)\citenamefont {Meng},
  \citenamefont {Feng}, \citenamefont {Liu}, \citenamefont {Wang},\ and\
  \citenamefont {Zou}}]{Meng:2021ecs}%
  \BibitemOpen
  \bibfield  {author} {\bibinfo {author} {\bibfnamefont {Y.}~\bibnamefont
  {Meng}}, \bibinfo {author} {\bibfnamefont {X.}~\bibnamefont {Feng}}, \bibinfo
  {author} {\bibfnamefont {C.}~\bibnamefont {Liu}}, \bibinfo {author}
  {\bibfnamefont {T.}~\bibnamefont {Wang}},\ and\ \bibinfo {author}
  {\bibfnamefont {Z.}~\bibnamefont {Zou}},\ }\href
  {https://doi.org/10.1016/j.scib.2023.07.041} {\bibfield  {journal} {\bibinfo
  {journal} {Sci. Bull.}\ }\textbf {\bibinfo {volume} {68}},\ \bibinfo {pages}
  {1880} (\bibinfo {year} {2023})}\BibitemShut {NoStop}%
\bibitem [{\citenamefont {Navas}\ \emph {et~al.}(2024)\citenamefont {Navas}
  \emph {et~al.}}]{ParticleDataGroup:2024cfk}%
  \BibitemOpen
  \bibfield  {author} {\bibinfo {author} {\bibfnamefont {S.}~\bibnamefont
  {Navas}} \emph {et~al.} (\bibinfo {collaboration} {Particle Data Group}),\
  }\href {https://doi.org/10.1103/PhysRevD.110.030001} {\bibfield  {journal}
  {\bibinfo  {journal} {Phys. Rev. D}\ }\textbf {\bibinfo {volume} {110}},\
  \bibinfo {pages} {030001} (\bibinfo {year} {2024})}\BibitemShut {NoStop}%
\bibitem [{\citenamefont {Adachi}\ \emph {et~al.}(2008)\citenamefont {Adachi}
  \emph {et~al.}}]{Belle:2006mzv}%
  \BibitemOpen
  \bibfield  {author} {\bibinfo {author} {\bibfnamefont {I.}~\bibnamefont
  {Adachi}} \emph {et~al.} (\bibinfo {collaboration} {Belle Collaboration}),\
  }\href {https://doi.org/10.1016/j.physletb.2008.03.029} {\bibfield  {journal}
  {\bibinfo  {journal} {Phys. Lett. B}\ }\textbf {\bibinfo {volume} {662}},\
  \bibinfo {pages} {323} (\bibinfo {year} {2008})}\BibitemShut {NoStop}%
\bibitem [{\citenamefont {Baglin}\ \emph {et~al.}(1987)\citenamefont {Baglin}
  \emph
  {et~al.}}]{AnnecyLAPP-CERN-Genoa-Lyon-Oslo-Rome-Strasbourg-Turin:1987qkd}%
  \BibitemOpen
  \bibfield  {author} {\bibinfo {author} {\bibfnamefont {C.}~\bibnamefont
  {Baglin}} \emph {et~al.} (\bibinfo {collaboration}
  {Annecy(LAPP)-CERN-Genoa-Lyon-Oslo-Rome-Strasbourg-Turin}),\ }\href
  {https://doi.org/10.1016/0370-2693(87)90098-0} {\bibfield  {journal}
  {\bibinfo  {journal} {Phys. Lett. B}\ }\textbf {\bibinfo {volume} {187}},\
  \bibinfo {pages} {191} (\bibinfo {year} {1987})}\BibitemShut {NoStop}%
\bibitem [{\citenamefont {Armstrong}\ \emph {et~al.}(1995)\citenamefont
  {Armstrong} \emph {et~al.}}]{E760:1995rep}%
  \BibitemOpen
  \bibfield  {author} {\bibinfo {author} {\bibfnamefont {T.~A.}\ \bibnamefont
  {Armstrong}} \emph {et~al.} (\bibinfo {collaboration} {E760 Collaboration}),\
  }\href {https://doi.org/10.1103/PhysRevD.52.4839} {\bibfield  {journal}
  {\bibinfo  {journal} {Phys. Rev. D}\ }\textbf {\bibinfo {volume} {52}},\
  \bibinfo {pages} {4839} (\bibinfo {year} {1995})}\BibitemShut {NoStop}%
\bibitem [{\citenamefont {Ambrogiani}\ \emph {et~al.}(2003)\citenamefont
  {Ambrogiani} \emph {et~al.}}]{FermilabE835:2003ula}%
  \BibitemOpen
  \bibfield  {author} {\bibinfo {author} {\bibfnamefont {M.}~\bibnamefont
  {Ambrogiani}} \emph {et~al.} (\bibinfo {collaboration} {Fermilab E835
  Collaboration}),\ }\href {https://doi.org/10.1016/S0370-2693(03)00805-0}
  {\bibfield  {journal} {\bibinfo  {journal} {Phys. Lett. B}\ }\textbf
  {\bibinfo {volume} {566}},\ \bibinfo {pages} {45} (\bibinfo {year}
  {2003})}\BibitemShut {NoStop}%
\bibitem [{\citenamefont {Adams}\ \emph {et~al.}(2008)\citenamefont {Adams}
  \emph {et~al.}}]{CLEO:2008qfy}%
  \BibitemOpen
  \bibfield  {author} {\bibinfo {author} {\bibfnamefont {G.~S.}\ \bibnamefont
  {Adams}} \emph {et~al.} (\bibinfo {collaboration} {CLEO Collaboration}),\
  }\href {https://doi.org/10.1103/PhysRevLett.101.101801} {\bibfield  {journal}
  {\bibinfo  {journal} {Phys. Rev. Lett.}\ }\textbf {\bibinfo {volume} {101}},\
  \bibinfo {pages} {101801} (\bibinfo {year} {2008})}\BibitemShut {NoStop}%
\bibitem [{\citenamefont {Ablikim}\ \emph {et~al.}(2013)\citenamefont {Ablikim}
  \emph {et~al.}}]{BESIII:2012lxx}%
  \BibitemOpen
  \bibfield  {author} {\bibinfo {author} {\bibfnamefont {M.}~\bibnamefont
  {Ablikim}} \emph {et~al.} (\bibinfo {collaboration} {BESIII Collaboration}),\
  }\href {https://doi.org/10.1103/PhysRevD.87.032003} {\bibfield  {journal}
  {\bibinfo  {journal} {Phys. Rev. D}\ }\textbf {\bibinfo {volume} {87}},\
  \bibinfo {pages} {032003} (\bibinfo {year} {2013})}\BibitemShut {NoStop}%
\bibitem [{\citenamefont {Ablikim}\ \emph {et~al.}(2025)\citenamefont {Ablikim}
  \emph {et~al.}}]{BESIII:2024rex}%
  \BibitemOpen
  \bibfield  {author} {\bibinfo {author} {\bibfnamefont {M.}~\bibnamefont
  {Ablikim}} \emph {et~al.} (\bibinfo {collaboration} {BESIII Collaboration}),\
  }\href {https://doi.org/10.1103/PhysRevLett.134.181901} {\bibfield  {journal}
  {\bibinfo  {journal} {Phys. Rev. Lett.}\ }\textbf {\bibinfo {volume} {134}},\
  \bibinfo {pages} {181901} (\bibinfo {year} {2025})}\BibitemShut {NoStop}%
\bibitem [{\citenamefont {Ablikim}\ \emph {et~al.}(2012)\citenamefont {Ablikim}
  \emph {et~al.}}]{BESIII:2012urf}%
  \BibitemOpen
  \bibfield  {author} {\bibinfo {author} {\bibfnamefont {M.}~\bibnamefont
  {Ablikim}} \emph {et~al.} (\bibinfo {collaboration} {BESIII Collaboration}),\
  }\bibfield  {title} {\bibinfo {title} {{Study of $\psi(3686)\to\pi^0 h_c,
  h_c\to\gamma\eta_c$ via $\eta_c$ exclusive decays}},\ }\href
  {https://doi.org/10.1103/PhysRevD.86.092009} {\bibfield  {journal} {\bibinfo
  {journal} {Phys. Rev. D}\ }\textbf {\bibinfo {volume} {86}},\ \bibinfo
  {pages} {092009} (\bibinfo {year} {2012})},\ \Eprint
  {https://arxiv.org/abs/1209.4963} {arXiv:1209.4963 [hep-ex]} \BibitemShut
  {NoStop}%
\bibitem [{\citenamefont {Ablikim}\ \emph {et~al.}(2018)\citenamefont {Ablikim}
  \emph {et~al.}}]{BESIII:2017tvm}%
  \BibitemOpen
  \bibfield  {author} {\bibinfo {author} {\bibfnamefont {M.}~\bibnamefont
  {Ablikim}} \emph {et~al.} (\bibinfo {collaboration} {BESIII Collaboration}),\
  }\href {https://doi.org/10.1088/1674-1137/42/2/023001} {\bibfield  {journal}
  {\bibinfo  {journal} {Chin. Phys. C}\ }\textbf {\bibinfo {volume} {42}},\
  \bibinfo {pages} {023001} (\bibinfo {year} {2018})}\BibitemShut {NoStop}%
\bibitem [{\citenamefont {Ablikim}\ \emph
  {et~al.}(2024{\natexlab{a}})\citenamefont {Ablikim} \emph
  {et~al.}}]{BESIII:2024lks}%
  \BibitemOpen
  \bibfield  {author} {\bibinfo {author} {\bibfnamefont {M.}~\bibnamefont
  {Ablikim}} \emph {et~al.} (\bibinfo {collaboration} {BESIII Collaboration}),\
  }\href {https://doi.org/10.1088/1674-1137/ad595b} {\bibfield  {journal}
  {\bibinfo  {journal} {Chin. Phys. C}\ }\textbf {\bibinfo {volume} {48}},\
  \bibinfo {pages} {093001} (\bibinfo {year} {2024}{\natexlab{a}})}\BibitemShut
  {NoStop}%
\bibitem [{\citenamefont {Ablikim}\ \emph {et~al.}(2010)\citenamefont {Ablikim}
  \emph {et~al.}}]{BESIII:2009fln}%
  \BibitemOpen
  \bibfield  {author} {\bibinfo {author} {\bibfnamefont {M.}~\bibnamefont
  {Ablikim}} \emph {et~al.} (\bibinfo {collaboration} {BESIII Collaboration}),\
  }\href {https://doi.org/10.1016/j.nima.2009.12.050} {\bibfield  {journal}
  {\bibinfo  {journal} {Nucl. Instrum. Meth. A}\ }\textbf {\bibinfo {volume}
  {614}},\ \bibinfo {pages} {345} (\bibinfo {year} {2010})}\BibitemShut
  {NoStop}%
\bibitem [{\citenamefont {Yu}\ \emph {et~al.}(2016)\citenamefont {Yu} \emph
  {et~al.}}]{Yu:2016cof}%
  \BibitemOpen
  \bibfield  {author} {\bibinfo {author} {\bibfnamefont {C.}~\bibnamefont {Yu}}
  \emph {et~al.},\ }in\ \href {https://doi.org/10.18429/JACoW-IPAC2016-TUYA01}
  {\emph {\bibinfo {booktitle} {{7th International Particle Accelerator
  Conference}}}}\ (\bibinfo {year} {2016})\ p.\ \bibinfo {pages}
  {TUYA01}\BibitemShut {NoStop}%
\bibitem [{\citenamefont {Agostinelli}\ \emph {et~al.}(2003)\citenamefont
  {Agostinelli} \emph {et~al.}}]{GEANT4:2002zbu}%
  \BibitemOpen
  \bibfield  {author} {\bibinfo {author} {\bibfnamefont {S.}~\bibnamefont
  {Agostinelli}} \emph {et~al.} (\bibinfo {collaboration} {GEANT4
  Collaboration}),\ }\href {https://doi.org/10.1016/S0168-9002(03)01368-8}
  {\bibfield  {journal} {\bibinfo  {journal} {Nucl. Instrum. Meth. A}\ }\textbf
  {\bibinfo {volume} {506}},\ \bibinfo {pages} {250} (\bibinfo {year}
  {2003})}\BibitemShut {NoStop}%
\bibitem [{\citenamefont {Allison}\ \emph {et~al.}(2006)\citenamefont {Allison}
  \emph {et~al.}}]{Allison:2006ve}%
  \BibitemOpen
  \bibfield  {author} {\bibinfo {author} {\bibfnamefont {J.}~\bibnamefont
  {Allison}} \emph {et~al.},\ }\href {https://doi.org/10.1109/TNS.2006.869826}
  {\bibfield  {journal} {\bibinfo  {journal} {IEEE Trans. Nucl. Sci.}\ }\textbf
  {\bibinfo {volume} {53}},\ \bibinfo {pages} {270} (\bibinfo {year}
  {2006})}\BibitemShut {NoStop}%
\bibitem [{\citenamefont {Allison}\ \emph {et~al.}(2016)\citenamefont {Allison}
  \emph {et~al.}}]{Allison:2016lfl}%
  \BibitemOpen
  \bibfield  {author} {\bibinfo {author} {\bibfnamefont {J.}~\bibnamefont
  {Allison}} \emph {et~al.},\ }\href
  {https://doi.org/10.1016/j.nima.2016.06.125} {\bibfield  {journal} {\bibinfo
  {journal} {Nucl. Instrum. Meth. A}\ }\textbf {\bibinfo {volume} {835}},\
  \bibinfo {pages} {186} (\bibinfo {year} {2016})}\BibitemShut {NoStop}%
\bibitem [{\citenamefont {Deng}\ \emph {et~al.}(2006)\citenamefont {Deng} \emph
  {et~al.}}]{2006ObjectorientedBD}%
  \BibitemOpen
  \bibfield  {author} {\bibinfo {author} {\bibfnamefont {Z.~Y.}\ \bibnamefont
  {Deng}} \emph {et~al.},\ }\href
  {https://api.semanticscholar.org/CorpusID:124741001} {\bibfield  {journal}
  {\bibinfo  {journal} {High Energy Physics and Nuclear Physics}\ }\textbf
  {\bibinfo {volume} {30}} (\bibinfo {year} {2006})}\BibitemShut {NoStop}%
\bibitem [{\citenamefont {Ya-Jun}\ \emph {et~al.}(2008)\citenamefont {Ya-Jun},
  \citenamefont {Yu-Tie},\ and\ \citenamefont {Zheng-Yun}}]{Ya-Jun:2008mrr}%
  \BibitemOpen
  \bibfield  {author} {\bibinfo {author} {\bibfnamefont {M.}~\bibnamefont
  {Ya-Jun}}, \bibinfo {author} {\bibfnamefont {L.}~\bibnamefont {Yu-Tie}},\
  and\ \bibinfo {author} {\bibfnamefont {Y.}~\bibnamefont {Zheng-Yun}},\ }\href
  {https://doi.org/10.1088/1674-1137/32/7/012} {\bibfield  {journal} {\bibinfo
  {journal} {Chin. Phys. C C}\ }\textbf {\bibinfo {volume} {32}},\ \bibinfo
  {pages} {572} (\bibinfo {year} {2008})}\BibitemShut {NoStop}%
\bibitem [{\citenamefont {Liang}\ \emph {et~al.}(2009)\citenamefont {Liang}
  \emph {et~al.}}]{Liang:2009zzb}%
  \BibitemOpen
  \bibfield  {author} {\bibinfo {author} {\bibfnamefont {Y.-T.}\ \bibnamefont
  {Liang}} \emph {et~al.},\ }\href {https://doi.org/10.1016/j.nima.2009.02.036}
  {\bibfield  {journal} {\bibinfo  {journal} {Nucl. Instrum. Meth. A}\ }\textbf
  {\bibinfo {volume} {603}},\ \bibinfo {pages} {325} (\bibinfo {year}
  {2009})}\BibitemShut {NoStop}%
\bibitem [{\citenamefont {Huang}\ \emph {et~al.}(2022)\citenamefont {Huang},
  \citenamefont {Li}, \citenamefont {Qian}, \citenamefont {Zhu}, \citenamefont
  {Li}, \citenamefont {Zhang}, \citenamefont {Sun},\ and\ \citenamefont
  {You}}]{Huang:2022wuo}%
  \BibitemOpen
  \bibfield  {author} {\bibinfo {author} {\bibfnamefont {K.-X.}\ \bibnamefont
  {Huang}}, \bibinfo {author} {\bibfnamefont {Z.-J.}\ \bibnamefont {Li}},
  \bibinfo {author} {\bibfnamefont {Z.}~\bibnamefont {Qian}}, \bibinfo {author}
  {\bibfnamefont {J.}~\bibnamefont {Zhu}}, \bibinfo {author} {\bibfnamefont
  {H.-Y.}\ \bibnamefont {Li}}, \bibinfo {author} {\bibfnamefont {Y.-M.}\
  \bibnamefont {Zhang}}, \bibinfo {author} {\bibfnamefont {S.-S.}\ \bibnamefont
  {Sun}},\ and\ \bibinfo {author} {\bibfnamefont {Z.-Y.}\ \bibnamefont {You}},\
  }\href {https://doi.org/10.1007/s41365-022-01133-8} {\bibfield  {journal}
  {\bibinfo  {journal} {Nucl. Sci. Tech.}\ }\textbf {\bibinfo {volume} {33}},\
  \bibinfo {pages} {142} (\bibinfo {year} {2022})}\BibitemShut {NoStop}%
\bibitem [{\citenamefont {Ablikim}\ \emph
  {et~al.}(2024{\natexlab{b}})\citenamefont {Ablikim} \emph
  {et~al.}}]{BESIII:2024hdv}%
  \BibitemOpen
  \bibfield  {author} {\bibinfo {author} {\bibfnamefont {M.}~\bibnamefont
  {Ablikim}} \emph {et~al.} (\bibinfo {collaboration} {BESIII Collaboration}),\
  }\href {https://doi.org/10.1103/PhysRevD.110.L031101} {\bibfield  {journal}
  {\bibinfo  {journal} {Phys. Rev. D}\ }\textbf {\bibinfo {volume} {110}},\
  \bibinfo {pages} {L031101} (\bibinfo {year}
  {2024}{\natexlab{b}})}\BibitemShut {NoStop}%
\bibitem [{\citenamefont {Barlow}(2002)}]{Barlow:2002yb}%
  \BibitemOpen
  \bibfield  {author} {\bibinfo {author} {\bibfnamefont {R.}~\bibnamefont
  {Barlow}},\ }in\ \href@noop {} {\emph {\bibinfo {booktitle} {{Conference on
  Advanced Statistical Techniques in Particle Physics}}}}\ (\bibinfo {year}
  {2002})\ pp.\ \bibinfo {pages} {134--144}\BibitemShut {NoStop}%
\bibitem [{\citenamefont {Ablikim}\ \emph {et~al.}(2022)\citenamefont {Ablikim}
  \emph {et~al.}}]{BESIII:2022jde}%
  \BibitemOpen
  \bibfield  {author} {\bibinfo {author} {\bibfnamefont {M.}~\bibnamefont
  {Ablikim}} \emph {et~al.} (\bibinfo {collaboration} {BESIII Collaboration}),\
  }\href {https://doi.org/10.1103/PhysRevD.106.112002} {\bibfield  {journal}
  {\bibinfo  {journal} {Phys. Rev. D}\ }\textbf {\bibinfo {volume} {106}},\
  \bibinfo {pages} {112002} (\bibinfo {year} {2022})}\BibitemShut {NoStop}%
\bibitem [{\citenamefont {Chen}\ \emph
  {et~al.}(2020{\natexlab{b}})\citenamefont {Chen}, \citenamefont {Gong},
  \citenamefont {Li}, \citenamefont {Liu}, \citenamefont {Liu}, \citenamefont
  {Liu}, \citenamefont {Ma}, \citenamefont {Meng}, \citenamefont {Xiong},\ and\
  \citenamefont {Zhang}}]{Chen_2020}%
  \BibitemOpen
  \bibfield  {author} {\bibinfo {author} {\bibfnamefont {Y.}~\bibnamefont
  {Chen}}, \bibinfo {author} {\bibfnamefont {M.}~\bibnamefont {Gong}}, \bibinfo
  {author} {\bibfnamefont {N.}~\bibnamefont {Li}}, \bibinfo {author}
  {\bibfnamefont {C.}~\bibnamefont {Liu}}, \bibinfo {author} {\bibfnamefont
  {Y.-B.}\ \bibnamefont {Liu}}, \bibinfo {author} {\bibfnamefont
  {Z.}~\bibnamefont {Liu}}, \bibinfo {author} {\bibfnamefont {J.-P.}\
  \bibnamefont {Ma}}, \bibinfo {author} {\bibfnamefont {Y.}~\bibnamefont
  {Meng}}, \bibinfo {author} {\bibfnamefont {C.}~\bibnamefont {Xiong}},\ and\
  \bibinfo {author} {\bibfnamefont {K.-L.}\ \bibnamefont {Zhang}},\ }\href
  {https://doi.org/10.1088/1674-1137/44/8/083108} {\bibfield  {journal}
  {\bibinfo  {journal} {Chinese Physics C}\ }\textbf {\bibinfo {volume} {44}},\
  \bibinfo {pages} {083108} (\bibinfo {year} {2020}{\natexlab{b}})}\BibitemShut
  {NoStop}%
\bibitem [{\citenamefont {Wang}\ \emph {et~al.}(2025)\citenamefont {Wang},
  \citenamefont {Ren}, \citenamefont {Shen}, \citenamefont {Wu}, \citenamefont
  {Di~Giustino},\ and\ \citenamefont {Brodsky}}]{Wang_2025}%
  \BibitemOpen
  \bibfield  {author} {\bibinfo {author} {\bibfnamefont {S.-Q.}\ \bibnamefont
  {Wang}}, \bibinfo {author} {\bibfnamefont {Z.-Y.}\ \bibnamefont {Ren}},
  \bibinfo {author} {\bibfnamefont {J.-M.}\ \bibnamefont {Shen}}, \bibinfo
  {author} {\bibfnamefont {X.-G.}\ \bibnamefont {Wu}}, \bibinfo {author}
  {\bibfnamefont {L.}~\bibnamefont {Di~Giustino}},\ and\ \bibinfo {author}
  {\bibfnamefont {S.~J.}\ \bibnamefont {Brodsky}},\ }\bibfield  {journal}
  {\bibinfo  {journal} {Physical Review D}\ }\textbf {\bibinfo {volume}
  {111}},\ \href {https://doi.org/10.1103/physrevd.111.094016}
  {10.1103/physrevd.111.094016} (\bibinfo {year} {2025})\BibitemShut {NoStop}%
\end{thebibliography}%

\end{document}